\documentclass[twocolumn]{aastex63}
\usepackage{xcolor}

\newcommand\etal{\it{et al.}\rm}
\def\MARU#1{{\ooalign{\hfil#1\/\hfil\crcr\raise.167ex\hbox{\mathhexbox20D}}}}
\accepted{April 27, 2021}
\submitjournal{PJAB}
\setcitestyle{numbers}

\shorttitle{Subaru Telesccope}
\shortauthors{Iye, M.}


\begin{document}

\title{Subaru Telescope\\
--- History, Active/Adaptive Optics, Instruments, and Scientific Achievements---}

\author{Masanori Iye}
\affil{National Astronomical Observatory of Japan, Osawa 2-21-1, Mitaka, Tokyo 181-8588 Japan}
%%National Astronomical Observatory of Japan, Osawa 2-21-1, Mitaka, Tokyo 181-8588 Japan
%\affil{National Institutes of Natural Sciences, Hulic Kamiyacho Building 4-3-13 Toranomon, Minato,  Tokyo 105-0001 Japan}

%% Note that the \and command from previous versions of AASTeX is now
%% depreciated in this version as it is no longer necessary. AASTeX 
%% automatically takes care of all commas and "and"s between authors names.

%% AASTeX 6.2 has the new \collaboration and \nocollaboration commands to
%% provide the collaboration status of a group of authors. These commands 
%% can be used either before or after the list of corresponding authors. The
%% argument for \collaboration is the collaboration identifier. Authors are
%% encouraged to surround collaboration identifiers with ()s. The 
%% \nocollaboration command takes no argument and exists to indicate that
%% the nearby authors are not part of surrounding collaborations.

%% Mark off the abstract in the ``abstract'' environment. 
\begin{abstract}
The Subaru Telescope\footnote{https://subarutelescope.org/en/}  is an 8.2 m optical/infrared telescope constructed during 1991--1999 and has been operational since 2000 on the summit area of Maunakea, Hawaii, by the National Astronomical Observatory of Japan (NAOJ). This paper reviews the history, key engineering issues, and selected scientific achievements of the Subaru Telescope. 
The active optics for a thin primary mirror was the design backbone of the telescope to deliver a high-imaging performance. Adaptive optics with a laser-facility to generate an artificial guide-star improved the telescope vision to its diffraction limit by cancelling any atmospheric turbulence effect in real time. Various observational instruments, especially the wide-field camera, have enabled unique observational studies.
Selected scientific topics include studies on cosmic reionization, weak/strong gravitational lensing, cosmological parameters, primordial black holes, the dynamical/chemical evolution/interactions of galaxies, neutron star mergers, supernovae, exoplanets, proto-planetary disks, and outliers of the solar system.
The last described are operational statistics, plans and a note concerning the culture-and-science issues in Hawaii.
\end{abstract}

%% Keywords should appear after the \end{abstract} command. 
%% See the online documentation for the full list of available subject
%% keywords and the rules for their use.
%%Keywords : Active Optics, Adaptive Optics, Telescope, Instruments, Cosmology, Exoplanets
%% From the front matter, we move on to the body of the paper.
%% Sections are demarcated by \section and \subsection, respectively.
%% Observe the use of the LaTeX \label
%% command after the \subsection to give a symbolic KEY to the
%% subsection for cross-referencing in a \ref command.
%% You can use LaTeX's \ref and \label commands to keep track of
%% cross-references to sections, equations, tables, \& figures.
%% That way, if you change the order of any elements, LaTeX will

%% automatically renumber them.
%%
%% We recommend that authors also use the natbib~\cite
%% \&~\citet commands to identify citations.  The citations are
%% tied to the reference list via symbolic KEYs. The KEY correspondsh
%% to the KEY in the \bibitem in the reference list below. 

%\chapter{Subaru Telescope project}

\section{\bf{Prehistory}}

\subsection{\bf{Okayama 188 cm Telescope}}
In 1953, Yusuke Hagiwara\footnote{Member of the Japan Academy in 1944--79}, director of the Tokyo Astronomical Observatory, University of Tokyo, emphasized in a lecture the importance of building a modern large telescope. He did so in the presence of the Showa Emperor and obtained a budget for the next year to build a telescope for astrophysical observations. This budget was the first case approved by the Japanese congress for a newly established budget scheme involving multiyear spending. The 188 cm telescope was manufactured by Grubb-Parsons in the UK and was installed in 1960 at the newly built Okayama Astrophysical Observatory. It was the largest telescope in Asia and the sixth largest in the world.  Jun Jugaku first identified a star with excess UV as an optical counterpart of the X-ray source Sco X-1~\cite{Sandage1966}.  Sco X-1 was an unknown X-ray source found at that time by observations using an X-ray collimator instrument invented by Minoru Oda\footnote{Member of the Japan Academy in 1988--2001}~\cite{Gursky1966}.  Yoshio Fujita\footnote{Member of the Japan Academy in 1965--2013 and the President of the Japan Academy during 1994--2000} studied the atmosphere of low-temperature stars. He and his school established a spectroscopic classification system of carbon stars~\cite{Fujita1970, Yamashita1972} using the Okayama 188cm telescope.

Among the three domestic sites evaluated, the Okayama Astrophysical Observatory site was considered to be the best for installing the 188 cm telescope. A fair percentage of clear nights, ca 30\%, and stable astronomical ``seeing" due to the moderate Seto Inland Sea climate were sufficiently good for stellar spectroscopy. However, the rapidly expanding Mizushima industrial complex increased night sky brightness. Observing faint objects, like galaxies, soon became increasingly difficult. The site was also not suitable for emerging infrared observations that require a high-altitude site for low precipitable water.

\subsection{\bf{Japan National Large Telescope Working Group}}
After completion of the 5 m Hale Telescope in 1948 at Palomar Mountain Observatory in California, a new generation of 4 m class telescopes were constructed during the 1970s and the 1980s. Apart from the 4 m Mayall Telescope built on Kitt Peak (1973)~\cite{Robinson1981} and the 3.9 m Anglo-Australian Telescope built on Siding Springs (1974)~\cite{Gascoigne1975}, many others were built on remote sites abroad, not limited to sites within the host country. These included the Blanco 4 m Telescope at Cerro Tololo (1974)~\cite{Edmondson1986}, European Southern Observatory(ESO) 3.6 m Telescope at La Silla (1977)~\cite{Blaauw1990}, 3.0 m NASA IRTF on Maunakea (1979)~\cite{Becklin1981}, 3.6 m Canada-France-Hawaii Telescope on Maunakea (1979)~\cite{Bely1982}, and 4.2 m William Herschel Telescope on La Palma (1987)~\cite{Boksenberg1985}.  

During the early 1980s, the Japanese optical astronomy community discussed how to establish a plan to build a new telescope. The initial plan was to build a 3.5 m telescope on a domestic site, since this appeared to be a natural step forward after establishing the Okayama 188 cm telescope. 
However, there were also voices to locate a new telescope on one of the best sites abroad. Enlarging the imagined telescope size to 7$\sim$8 m, while aiming at the world top class, was another point of discussion.  To make such a challenging jump, or to follow a secure path, was a topic of much debate. Having no experience for the funding ministry to build a national research facility abroad was another concern. However, the astronomical community decided to move ahead for a larger telescope by the end of 1984.

In August 1984, Keiichi Kodaira, the project leader, summoned a working group to start a design study of a targeted 7.5 m telescope. The telescope was tentatively named Japan National Large Telescope (JNLT), and was to be built on a great site abroad. This author chaired the working group and held 50 meetings with scientists and engineers from the academic community and industries. These meetings were held in 1984--89 to define the outline of the JNLT project.  

Yoshihide Kozai\footnote{Member of the Japan Academy in 1980--2018}, the director of the Tokyo Astronomical Observatory of the University of Tokyo, submitted a budget request; and the first preparatory study was authorized from 1987. The budget request for construction~\cite{Kodaira1989} was finally approved in 1990.

A known empirical relation indicated that the cost of building a telescope increases at the rate between the quadratic and cubic power of the diameter of the primary mirror. To the best of astronomical accuracy, roughly half of the total cost would be for the product components, and the remaining half would be for management. Therefore, to keep the cost at an affordable level, the first essential requirement was to reduce the mass of the primary mirror, around which the telescope structure is designed. Toward this goal, three separate approaches were under development in the mid-1980s to design and build a large light-weight primary mirror. 

\subsection{\bf{Defining the Telescope}}

\subsubsection{\bf{Choice of the Primary Mirror}}
\paragraph{\bf{Segmented Mirror}}

Jerry Nelson of California University announced a challenging plan to build a low-cost 10 m telescope with 54 segmented mirrors phased together to form a primary mirror~\cite{Nelson1979}.  The JNLT Working Group (WG) studied this innovative concept~\cite{Mast1985} during its early phase of the study. The WG found this approach to be too challenging to adopt without first having feasibility demonstrations, and dropped this option from further study.

The University of California and the California Institute of Technology, however, made a brave decision to adopt this plan~\cite{Nelson1982}. In 1985, they started construction of a 10 m telescope with the financial aid of Keck Foundation. 
The segmented mirrors were assembled in April 1992 and after years of effort to fine tune the system, an image quality of 0".42 (80$\%$ encircled energy) was attained to our surprise~\cite{Nelson1992, Wizinowich1994}. Thus, eventually, two 10 m Keck telescopes--Keck I in 1993 and Keck II in 1996--were completed on Maunakea.

Many engineering efforts and practical mathematics were behind the success of the segmented mirror technology. The development of edge sensors~\cite{Mast1983}, careful mathematical analyses~\cite{Mast1985}, and polishing the aspheric surface by bending mirrors~\cite{Mast1990} were all indispensable. This technology became a legacy for next-generation telescopes: Spanish 10.4 m telescope (GTC, 2009)~\cite{Rodriguez1997}, Thirty Meter Telescope (TMT)~\cite{Sanders2013}, and European Extremely Large Telescope (E-ELT)~\cite{Gilmozzi2007}. In fact, the original Spanish 8 m telescope proposal was made based on the thin meniscus mirror. However, Jerry Nelson, as an international reviewer, was successful to convince the team to adopt the segmented mirror technology.

The few lead years of the Keck Telescopes in starting scientific observations over the competing Very Large Telescope (VLT), Subaru Telescope, and Gemini Telescope, was a great advantage for California astronomers. The three national or international projects adopting thin meniscus mirrors referred to this situation with some sentiment of regret and respect that 8 m class telescopes were ``Kecked".

\paragraph{\bf{Honeycomb Mirror}}
Built in 1948, the 5 m Palomar telescope used another technology for its primary mirror. To reduce the glass mass, numerous clay cores were deployed in the melting furnace to produce a honeycomb sandwich glass structure. Roger Angel further developed this method by spin-casting the glass to form a parabolic surface to save on the  polishing time~\cite{Angel1982}. The melted glass was poured between the cores and a months-long careful cooling process followed so as to avoid the accumulation of internal stress in the glass. The clay cores were removed from the back and a monolithic honeycomb glass structure was completed with hollow spaces inside to make it lighter in weight.  Roger Angel further developed this technology in his Mirror Laboratory at the University of Arizona to produce light-weight honeycomb structure glass pieces and to polish these pieces into astronomical primary mirrors~\cite{Goble1986}. 

Borosilicate glass, e.g., Ohara E6, which has a low melting temperature and a low viscosity, was found to be suitable for casting in this furnace. However, the thermal expansion coefficient of borosilicate glass is on the order of $10^{-6}/^{\circ}C$, about 100-times more than those of so-called zero-expansion glasses. The JNLT WG spotted the need to develop the thermal control system for the primary mirror so as to avoid significant thermal deformation under varying ambient temperature during the night and to maintain a good imaging quality, which appeared to be another rather challenging issue~\cite{Angel1986}. 

A finite element analysis of honeycomb mirror deformation showed that although the thickness of the honeycomb mirror is a few-times larger than the thin meniscus mirror that was studied for a  comparison, the deformation would not be negligible, and that the active optics discussed in the following section would be required anyway. The WG concluded that force control is more manageable than thermal control.

The honeycomb mirror technology was adopted in many US university telescopes, for example, the 6.5 m Magellan 1 (2000) and 2 (2002) Telescopes~\cite{Shectman2003}, the Large Binocular Telescope with twin 8.3 m mirrors~\cite{Hill2006}, the Large Synoptic Survey Telescope~\cite{Tyson2001}, and the Giant Magellan Telescope with seven 8.4 m mirrors~\cite{Johns2012}. 

\paragraph{\bf{Thin Monolithic Mirror}}
Raymond Wilson at the ESO was the first promoter of the active optics concept: to use a thin monolithic mirror supported by numerous actuators so that the mirror shape can be actively controlled to maintain the optical surface with satisfying an optical specification~\cite{Wilson1987}. This author spent one year at ESO during 1983--84 and had early discussions on this concept with his group.  At that time, ESO was developing a 1 m test facility to show that the mirror surface shape could be controlled with supporting actuators~\cite{Noethe1986, Noethe1988}. For a monolithic mirror, one can adopt zero-expansion glass instead of borosilicate glass to ensure better thermal performance with a smaller internal stress. The force control of the reflective surface shape of a monolithic mirror is much more straightforward than controlling the phase of segmented separate mirrors. This approach was eventually adopted, not just for the 8.2 m Subaru Telescope, but also for the 8.1 m VLTs~\cite{Noethe1992, Giacconi1998}, and the 8.0 m Gemini Telescopes~\cite{Mountain1998}.

\vspace{5mm}
\section{\bf{Active optics}}
The basic idea of active optics is simple and mathematical.
Any departure of the mirror surface shape, $z(r,\theta)$, from the ideal hyperbolic shape required for  Ritchey-Chr\'{e}tien optics can be expanded in terms of Fourier-Zernike functions, which constitute one of the orthonormal sets of eigenfunctions of a circle, by the following expansion:

\begin{eqnarray}
\hspace{-2cm} z(r,\theta) = \sum^{\infty}_{n=2}A^{0}_{n}R^{0}_{n}(r) + \nonumber \\
 \sum^{\infty}_{n=1}\sum^{\infty}_{m=1}\bigl\{ (A^{m}_{n}\cos{m\theta} + B^{m}_{n} \sin{m\theta}) R^{m}_{n}(r) \bigr\},
\end{eqnarray}

\noindent where $R^{m}_{n}(r)$ is the radial function of the Zernike circle polynomials, $A^{m}_{n}$ and $B^{m}_{n}$ are the expansion coefficients with $n-m$ being even. 

The mirror shape error is therefore given by the vector $\bf{a}=\it\langle a_{i} \rangle(i=1,..,N)$, where the components $a_{i}$ are the coefficients $A^{m}_{n}$ and $B^{m}_{n}$ truncated to the first $N$ terms. Similarly, the correction force distribution is represented by another vector, $\bf{f} = \it\langle f_{j} \rangle (j=1,..,M)$, where the components $f_{j}$ are the correction forces of $M$ actuators.

The active optics uses Hooke's law of deformation through the following equation:

\begin{eqnarray}
& \bf{a}\it = C^{j}_{i} \bf{f}
\end{eqnarray}

\noindent where $C^{j}_{i}$ is the structure matrix of the mirror~\cite{Iye1990}.
This structure matrix, $C^{j}_{i}$, can be obtained by measuring the mirror deformation response to the unit force exerted for each of the actuators.  After the structure matrix is known, the correction force distribution for a measured deformation of the mirror surface can be derived by

\begin{eqnarray}
\bf{f}\it = -[C^{j}_{i}]^{-1} \bf{a}.
\end{eqnarray}

\subsection{\bf{Active Optics R{\&}Ds}}

\begin{figure}[h]
\centering
\includegraphics[scale=0.45]{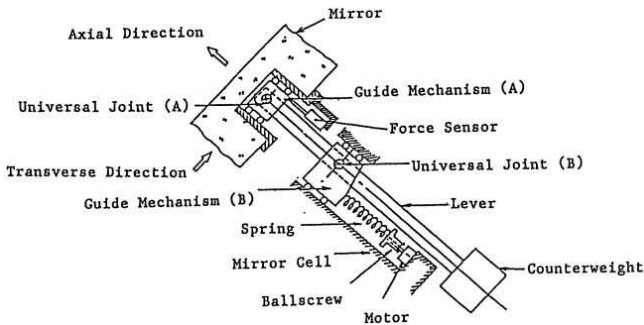}
\vspace{1mm} \caption{Structure of the electromechanical actuator developed for Subaru Telescope~\cite{Iye1989}.}
\label{actuator}
\end{figure}

Figure \ref{actuator} shows a schematic drawing of the electromechanical actuator developed for the JNLT active optics system. A force sensor developed by Shinko Denshi, which measures the frequency modulation of an elinvar-alloy tuning fork under tension, enabled force measurements with a resolution of 0.01 N over a 0--1500 N range. This force sensor was integrated in the actuator to control the supporting force with a $10^{-5}$ resolution at 1 Hz. 

\begin{figure}[h]
\centering
\includegraphics[scale=0.48]{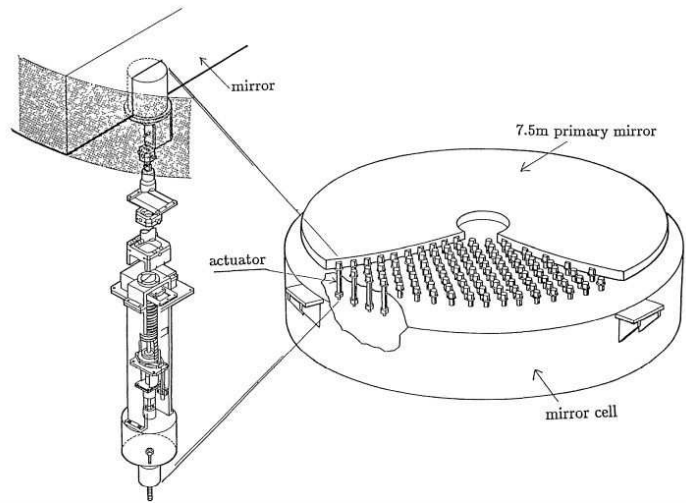}
\vspace{1mm} \caption{Deployment of 261 actuators to support the primary mirror of the Subaru Telescope~\cite{Iye1989}. The actuators, inserted from the back pocket holes, support the mirror at the local center of its gravity plane, actively in the axial direction and passively in the lateral direction.}
\label{MirrorSupport}
\end{figure}

\begin{figure}[h]
\centering
\includegraphics[scale=0.48]{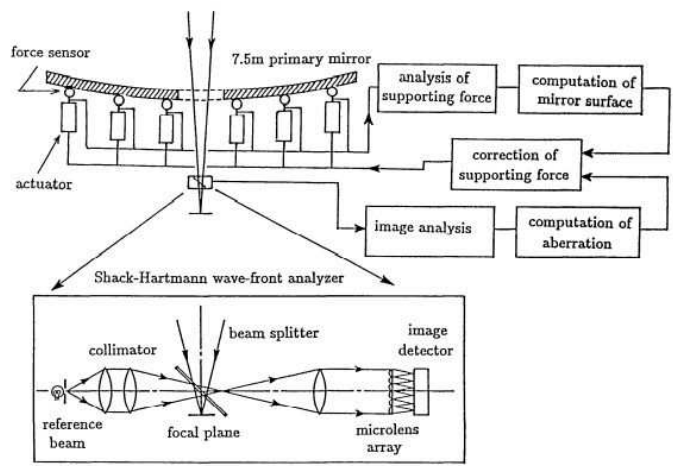}
\vspace{1mm} \caption{Control loop of the Subaru Telescope active optics system~\cite{Iye1989}.}
\label{ActiveOpticsLoop}
\end{figure}

Figure \ref{MirrorSupport} illustrates the deployment of 261 actuators to support the primary mirror at its local center of gravity, designed by Noboru Itoh, the lead engineer of MITSUBISHI ELECTRIC Co. The actuators work actively in the axial support and passively for the lateral mirror support. This system, hence, avoids any S-distortion~\cite{Schwesinger1969}, which arises for the mirrors supported laterally at its periphery during tilted orientations.  Figure \ref{ActiveOpticsLoop} shows the force control loop of JNLT, which is calibrated by optical Shack-Hartman camera measurements every time the primary mirror undergoes some operation including realuminization~\cite{Iye1989}. Owing to its stable configuration it was proven that there is no need to make nightly calibrationS of the structure matrix.

%~\citet{Iye1992} 

\begin{figure}[h]
\centering
\includegraphics[scale=0.3]{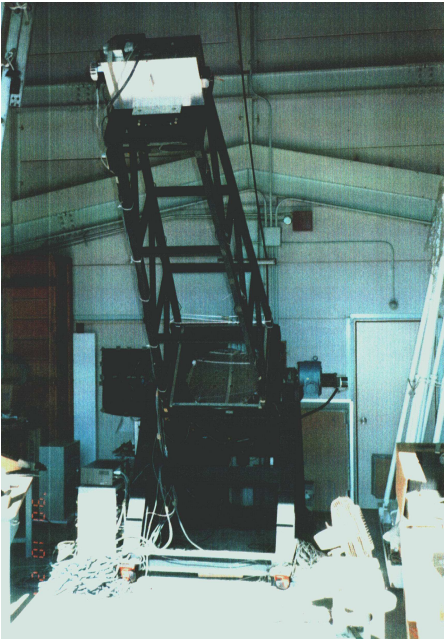}
\vspace{1mm} \caption{(Color online) Active optics experiment setup with a 62 cm spherical mirror supported by nine actuators and three fixed points and the Shack-Hartmann camera mounted on the top of the telescope structure.}
\label{62cm telescope}
\end{figure}

Although the success of controlling the 1 m test mirror shape in a vertical position~\cite{Noethe1986} was encouraging news, JNLT WG realized the need to conduct an independent verifying experiment of active optics with an original facility at tilted orientations. In 1988, the National Astronomical Observatory of Japan (NAOJ) reorganized from the Tokyo Astronomical Observatory (see section 3.4) and MITSUBISHI ELECTRIC Corporation (MELCO) developed an experimental telescope to verify that the active optics can control the mirror figure,  even at tilted orientations of the telescope (Figure \ref{62cm telescope}). A thin spherical mirror of 62 cm in diameter was fabricated and supported by nine prototype active optics actuators~\cite{Nishimura1988} developed by MELCO and by three fixed supporting points.  NAOJ developed a Shack-Hartmann camera~\cite{Noguchi1989} to measure the mirror shape from the top of the test telescope.

\begin{figure}[h]
%\centering
\includegraphics[scale=0.52]{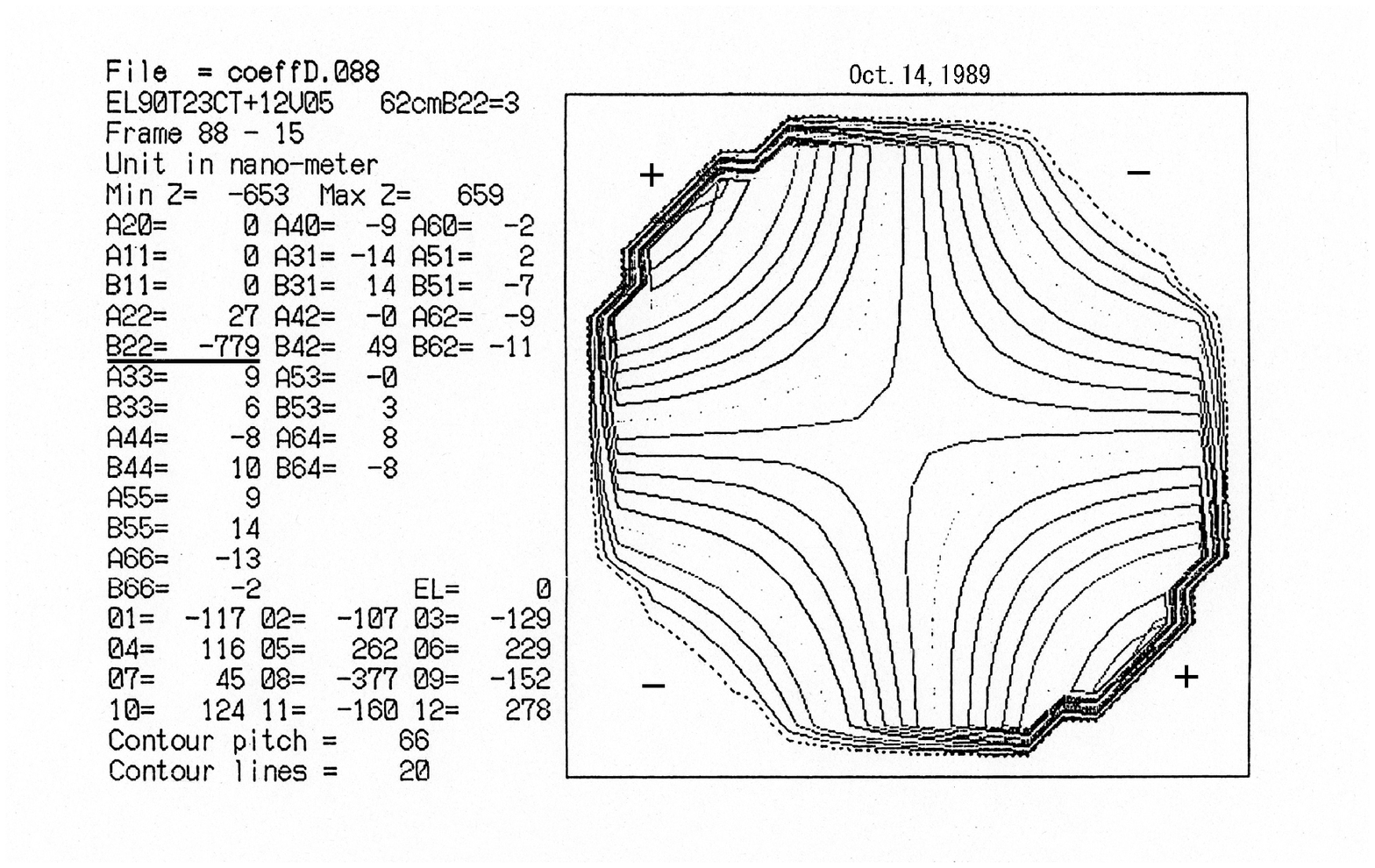}
\vspace{1mm} \caption{Measured result of the mirror shape control to generate an astigmatism deformation of $B_{22}$ = -779nm.}
\label{shape-control}
\end{figure}

Figure \ref{shape-control} shows the result of successful bending of the 62 cm spherical mirror to generate the specified amplitude of astigmatism $B^{2}_{2}$ = -780 nm, while keeping other terms at zero.  The measured results with this simple system were, $B^{2}_{2}$ = -779 nm, and the amplitude of other terms was less than 49 nm. This result ensured that one can correct  arbitrary lower-order Zernike terms of deformation of the primary mirror by implementing a larger number of actuators for JNLT.

%\begin{figure}[h]
%\centering
%\includegraphics[scale=0.45]{62cmsucc.eps}
%\vspace{1mm} \caption{A measured result of the mirror shape control to generate an astigmatism deformation of $A_{22}$=-779nm.}
%\label{shape-control}
%\end{figure}

\begin{figure}[h]
\centering
\includegraphics[scale=0.47]{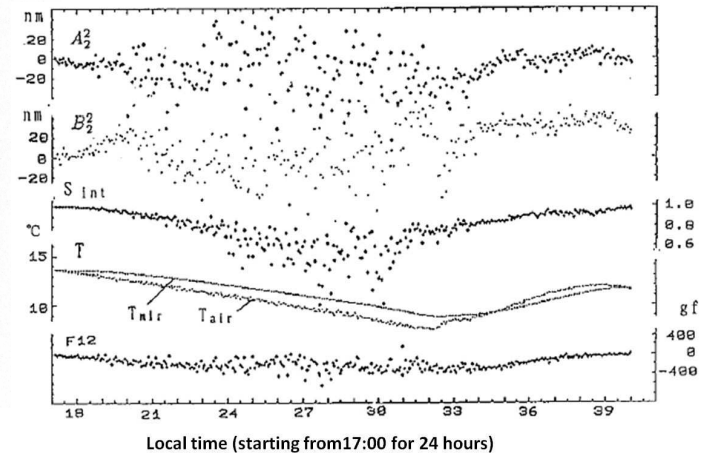}
\vspace{1mm} \caption{Time profiles of measurements of the astigmatic deformation amplitudes $A_{22}$ and $B_{22}$, the Strehl intensity $S_{int}$, the temperature of the ambient air $T_{air}$ and the mirror surface $T_{mir}$ are shown for every 3 min, starting from 1700 hours local time for 24 hours.  The measured values of the astigmatic deformation scatters during the period when the mirror surface is warmer than the air, $T_{mir}>T_{air}$, indicating that the microturbulence generated from the mirror surface disturbs the measurement, which is known as {\lq}{\lq}mirror seeing{\rq}{\rq}~\cite{Iye1990}.}
\label{mirrorseeing}
\end{figure}

Because the measuring sensitivity and stability of the mirror surface shape by the Shack-Hartmann sensor were critical for the active optics system, in 1989, automatic nighttime measurements of the mirror surface shape were conducted every three minutes at the NAOJ Mitaka campus to calibrate the measuring stability. Figure \ref{mirrorseeing} shows the measured amplitude profiles of astigmatism, $A^2_2$ and $B^2_2$, for 24 hours. This experiment was planned with an expectation that the nighttime measurements would provide cleaner data due to reduced car traffic near to the campus and the absence of human activity-related noises.  As is shown in Figure \ref{mirrorseeing}, contrary to WG's expectation, the nighttime measurement from around 1900 hours until around 1000 hours the next morning was more chaotic than that during daytime measurements.  

This phenomenon was interpreted as being due to microturbulence generated from the mirror surface. The mirror temperature drops belatedly than the cooling environmental air due to its thermal inertia. Therefore, the cooling mirror remains warmer than the ambient air during the nighttime. The on/off transition of the ``mirror seeing" is very well correlated with the temperature inversion between the air and the mirror, as shown in this figure. 

This finding suggests that all observations using existing ground-based telescopes were conducted through the ``mirror seeing" turbulence generated from their own mirrors. Based on this finding, a design provision was made for JNLT to keep the primary mirror at $2^{o}C$ below the foreseen nighttime temperature during the daytime by air conditioning the mirror inside the closed mirror cell.

The 62 cm experimental telescope setup was used to verify the active optics performance at tilted orientations~\cite{Iye1989b} and the effect of wind loading~\cite{Itoh1990}. Thus, adoptimg active optics for JNLT was decided.

\subsection{\bf{Seven Foci}}

The additional merit of adopting active optics is its flexibility in configuring the telescope foci. The primary mirror and the hyperbolic Cassegrain secondary mirror form Ritchey-Chr\'{e}tien optics optimized to provide a wide-field of view of 6 arcmin in diameter for the Cassegrain focus.  Due to the structural design, the distance from the intersection of the azimuthal axis and the elevation axis of the telescope to the Nasmyth foci is 1.6 m larger than the corresponding distance to the Cassegrain focus. Therefore, the Ritchey-Chr\'{e}tien optics for the Nasmyth focus should have a slightly different aspheric constant. Because infrared observations require a secondary mirror as an entrance pupil to shut out the thermal background, the size of the infrared secondary is smaller than the optical secondary mirror. The image at the Nasmyth foci rotates during exposure, and image rotators are equipped to de-rotate the image for the observing instruments.

Table \ref{table:JNLTOptics} summarizes the optical parameters for the seven foci: primary focus (F/1.87), optical and infrared Cassegrain foci (F/12.2, F/12.4), optical and infrared Nasmyth foci with dedicated image rotators (F/12.7 and F/13.9), and optical and infrared Nasmyth foci without an image rotator (F/12.5 and F/13.6).
JNLT, therefore, has three dedicated secondary mirrors: optical Cassegrain secondary (CasOptM2), optical Nasmyth secondary (NasOptM2), and infrared Cassegrain secondary (CasIRM2); for which the Ritchey-Chr\'{e}tien optics is configured, respectively, by 
changing the aspheric shape of the primary mirror. 

The infrared Nasmyth foci with and without the image rotator and the optical Nasmyth focus without the image rotator are configured without producing their dedicated secondary mirrors. However, the existing secondary mirrors are used to deliver a good image quality by adjusting the aspheric constant of the primary mirror and shifting the position of the secondary mirror to remove any spherical aberration.

The primary focus requires a dedicated field corrector lens system that also enables atmospheric dispersion corrections by shifting a lens unit laterally according to the elevation of the telescope~\cite{Nariai1994, Miyazaki2018}.

\begin{table}[h]
\begin{center}
\vspace{1mm} \caption{Optical parameters of the seven foci. Nasmyth Opt1 and Nasmyth IR1 are equipped with respective dedicated image rotator units. Nasmyth Opt2 and Nasmyth IR2 are not equipped with image rotators and do not constitute Ritchey-Chr\'{e}tien optics.}
\begin{tabular}{|c|c|c|c|} \hline
(1)&(2)&(3)&(4)\\
Focus&b1&b2&Top Unit\\ \hline
Prime&-1.0082456&NA&Corrector Lens\\
Cassegrain Opt&-1.0082456&-1.91608&CasOptM2\\
Cassegrain IR&-1.0084109&-1.91787&CasIRM2\\
Nasmyth Opt1&-1.0083945&-1.86543&NasOptM2\\
Nasmyth IR1&-1.019724&-1.91787&CasIRM2\\
Nasmyth Opt2&-1.0061814&-1.86543&NasOptM2\\
Nasmyth IR2&-1.017608&-1.91787&CasIRM2\\
\hline
\end{tabular}
\label{table:JNLTOptics}
\end{center}
\end{table}

\vspace{5mm}
\section{\bf{Project Promotion}}

\subsection{\bf{Site Survey}}
The JNLT WG made a worldwide site survey to identify the best site for building Subaru Telescope. Maunakea, Hawaii, Chilean sites on the Andes, and Canary Islands were studied with the support of Japan Society for the Promotion of Science (JSPS) for international studies during 1986--88. 
The group eventually chose Maunakea, Hawaii, as its most preferable site for construction owing to its superb observing conditions in the northern hemisphere. A high percentage of clear nights, dark sky, low water vapor, and superb seeing statistics were appreciated. A relatively shorter distance from Japan and cultural closeness with the presence of Japanese-American community in Hawaii were additional merits for the decision.

\subsection{\bf{Japan-UK Collaboration}}
Professor Richard Ellis, the chair of Large Telescope Panel of UK Science and Engineering Research Council, and other scientists visited Tokyo Astronomical Observatory in 1987 to discuss a potential Japan-UK equal partnership collaboration framework for building an 8 m class telescope.  Although this offer was appreciated, Tokyo Astronomical Observatory decided to decline the offer for two reasons.  First, JNLT will be the first national science project to build a facility abroad and building a national facility abroad is a high enough legal challenge, since there was no previous example.  Second, Tokyo Astronomical Observatory had no past experience to conduct this level of huge international collaboration.

Instead, Japan and the UK agreed to start collaboration on science programs and the development of instruments.  This author and Richard Ellis submitted proposals to JSPS and Particle Physics and Astronomy Research Council (PPARC), respectively, to enable this bi-lateral collaboration. A series of Japan-UK N+N meetings were established in 1993, and among others the mosaic charge-coupled device (CCD) camera developed by Maki Sekiguchi~\cite{Sekiguchi1992, Kashikawa1995} was brought to La Palma to conduct an imaging survey of clusters of galaxies in 1995--97 on the 4.2 m William Herschel Telescope~\cite{Komiyama2002}.

\subsection{\bf{Agreement with the University of Hawaii}}
In 1980, Yasumasa Yamashita, the division director of optical astronomy of Tokyo Astronomical Observatory, made the first contact with John Jefferies, the former director of the Institute for Astronomy (IfA) of the University of Hawaii, for a potential construction of the Japanese telescope facility on Maunakea. After exchanging letters and having meetings for several years, the Operation and Site Development Agreement (OSDA) was finally signed in 1988.  This agreement declared that NAOJ will try to secure funds to construct Subaru Telescope on Maunakea, while IfA would provide all of the necessary help to get it constructed and operated. In return for access to the summit ridge and a site for a Subaru base facility in Hilo with an annual nominal lease payment of US \$1 each, it was agreed that IfA would have guaranteed access for 52 observing nights a year. 

\subsection{\bf{Budget Request}}

Tokyo Astronomical Observatory submitted the budget request through University of Tokyo for preparatory studies of JNLT from 1987. In July 1988, Tokyo Astronomical Observatory became an interuniversity facility, the National Astronomical Observatory of Japan (NAOJ).  A full budget request was submitted in 1989. By that time, the JNLT concept was fixed to a 7.5 m active optics telescope to be built on Maunakea with four foci equipped to allow the installation of various imaging and spectroscopic instruments for optical and infrared observations. 

The astronomical community was in full support of this project and the Science Council of Japan (SCJ)  issued a strong recommendation to get it funded. The Ministry of Education, Science and Culture (MESC) started funding the construction of Subaru Telescope during 1991--99 as one of the eight major Frontier Science Projects approved.

\vspace{5mm}
\section{\bf{Construction}}

At the start of construction in 1991, the project team solicited a new name for JNLT and chose ``Subaru Telescope" after the ancient Japanese word for the Pleiades constellation. 

 The primary mirror was originally targeted for a size of 7.5 m in diameter (F/2.0) with a focal length of 15 m. It was enlarged in the early phase of construction to 8.2 m in diameter (F/1.83), which keeping the focal length of 15 m so as to minimize the design change to the enclosure and to the cost. This size change was decided by considering the announced size of 8.1 m VLT and the 8.0 m Gemini telescope. The aspheric constant for the Cassegrain Ritchey-Chr\'{e}tien system would be $b = -1.008350515$.

The Subaru Telescope base facility was completed in an astronomical complex of the University of Hawaii, Hilo campus, in 1996. After five years of constructing and installing facilities at the summit of Maunakea, the engineering first-light observation was successfully conducted on December 24, 1998.

\subsection{\bf{8 m-Club}}
In 1993, the Association of Universities for Research in Astronomy (AURA) and National Optical Astronomy Observatory (NOAO) of the United States decided to construct twin 8.0 m telescopes, Gemini, with one on Maunakea and the other on Cerro Tololo to cover both in the northern and southern skies. Their decision to employ the thin mirror active optics concept was challenged by some from the user community of honeycomb mirror telescopes. The Houck Committee in the US Congress summoned a hearing to confirm the appropriateness of adopting the thin mirror option.  
Upon a call from the Gemini group, a meeting with the VLT and Subaru representatives to discuss the merits and demerits of thin mirror projects was arranged in the UK. The meeting was named the ``8 m-club", and three more meetings were held within a year to discuss engineering issues, which are common or different among the three projects. 

The Subaru Telescope was unique to implement the prime focus for wide-field imaging observations. VLT and Gemini chose a design strategy to build multiple telescopes at low cost, four and two, respectively, so as to increase the total observation time. The higher construction cost of the Subaru Telescope was criticized during the budget request phase. The funding ministry, MEXT, however, approved the project. The choice of VLT and Gemini for lightweight telescopes made retrofitting a wide-field camera at the top of the telescope unfeasible because of their less sturdy structure and no space available to accommodate the  protruding camera inside the just-sized enclosure. Thus, Subaru Telescope became a unique 8--10 m class telescope, enabling wide-field imaging survey observations, which eventually turned out to be successful. 

Another concern expressed concerning the Subaru Telescope architecture was its challenging active support concept of drilling 261 pockets in the back of the primary mirror to support the mirror at its center-of-gravity plane. This design was also of an anxious concern for the Subaru project team, which considering an inherent risk of breakage of the precious primary mirror. The Subaru Telescope group, however, chose this design for obtaining better performance. Precautions were taken by removing microcracks at the ground pockets through acid etching and designing the support mechanism such that the exerted strain would not exceed 1000 psi. The team considered that this design with additional care concerning thermal control to reduce atmospheric turbulence in and around the telescope would help to attain a superb image quality.

\subsection{\bf{Telescope Structure}}
The telescope structure was designed to deliver a good image quality to four foci.  Keeping the optical alignment at any orientation in a varying thermal environment was a fundamental requirement. The telescope drive should ensure a high pointing and tracking accuracy for stable long-time exposure. To meet these scientific requirements, the error budget to attain the targeted performance was developed. Hydrostatic bearings and a direct drive were employed to drive the telescope. The finite-element method was widely applied to the full model of the telescope so as to evaluate its static and dynamic performance and, subsequently, improve the actual design of the telescope structure~\cite{Miyawaki1994}.

\begin{figure}[h]
\centering
\includegraphics[scale=0.45]{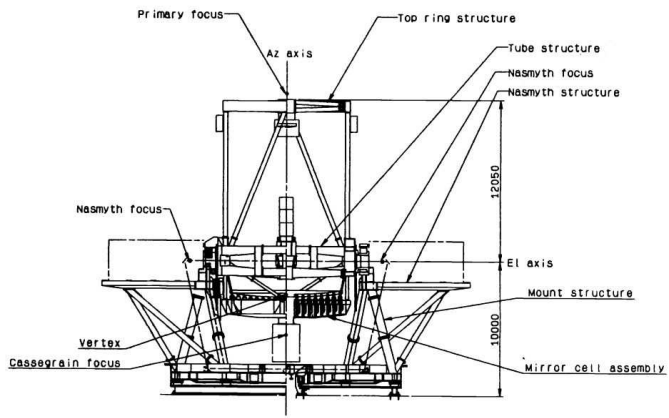}
\vspace{1mm} \caption{Side view of the Subaru Telescope structure (NAOJ).}
\label{TelescopeStructure}
\end{figure}

\begin{figure}[h]
\centering
\includegraphics[scale=0.3]{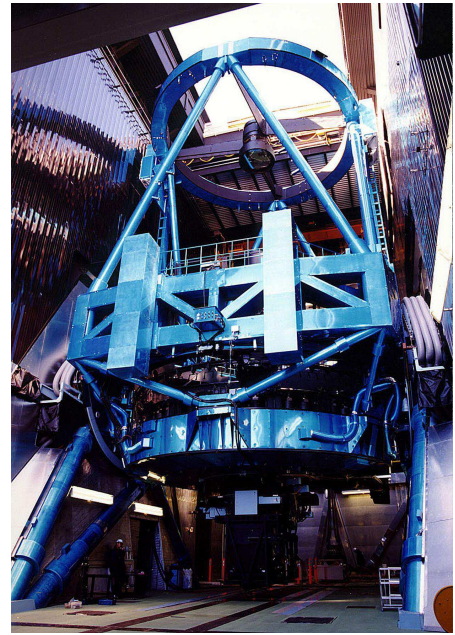}
\vspace{1mm} \caption{(Color online) Subaru Telescope installed in the enclosure. Black-painted ``Great Walls" on both sides of the telescope structure introduces laminar wind flow through the telescope so as to reduce the ``dome seeing" turbulence (NAOJ).}
\label{Telescope}
\end{figure}

The final structure of the telescope has a 22.2 m height and 27.2 m width with a total moving mass of 555 tons. 
Figure \ref{TelescopeStructure} illustrates the outline of the telescope structure. 
The telescope structure was once assembled and tested concerning its drive system in Japan in 1996, and then disassembled for shipment to Hawaii for the final installation. Figure \ref{Telescope} shows the actual telescope structure installed in the enclosure.

\subsection{\bf{Mirror Blank}}

\begin{figure}[h]
\centering
\includegraphics[scale=0.3]{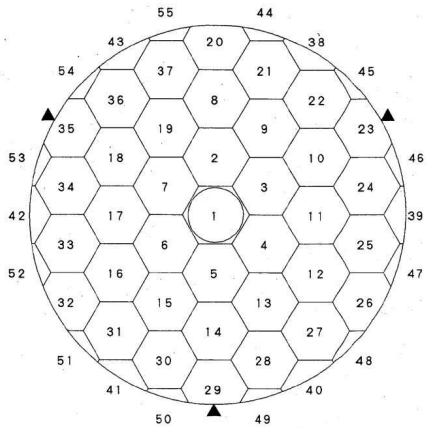}
\vspace{1mm} \caption{Hex disk deployment configuration for fusing into a 8.3 m monolithic mirror blank ~\cite{Nishiguchi1995}.}
\label{boules}
\end{figure}

Corning Glass Work produced 44 disks of 2 m in diameter, made of Ultra Low Expansion (ULE) glass, for the Subaru primary mirror blank. These disks were sliced in a hexagonal shape or smaller pieces, and their coefficients of thermal expansion (CTE), a few ppbs/$^{\circ}C$, were carefully measured. These pieces were placed in a fusing furnace to form a monolithic disk of 8.3 m in diameter and 30 cm thickness (Figure \ref{boules}). As the 31 hexagonal segments, each having a small CTE, but not exactly zero, are of the same shape and interchangeable, one should be able to find the best deployment configuration so that the resulting monolithic disk has the best thermal deformation behavior. A simulated annealing method was used to find the best deployment configuration among ca 100,000 patterns evaluated~\cite{Mikami1992, Sasaki1994}. This optimization reduced the thermal deformation amplitude of the final mirror to be $\sim1/30$ of those for a randomly arranged distribution. This study helped to attain a better image quality.

\subsection{\bf{Mirror Polishing}}

\begin{figure}[h]
\centering
\includegraphics[scale=0.5]{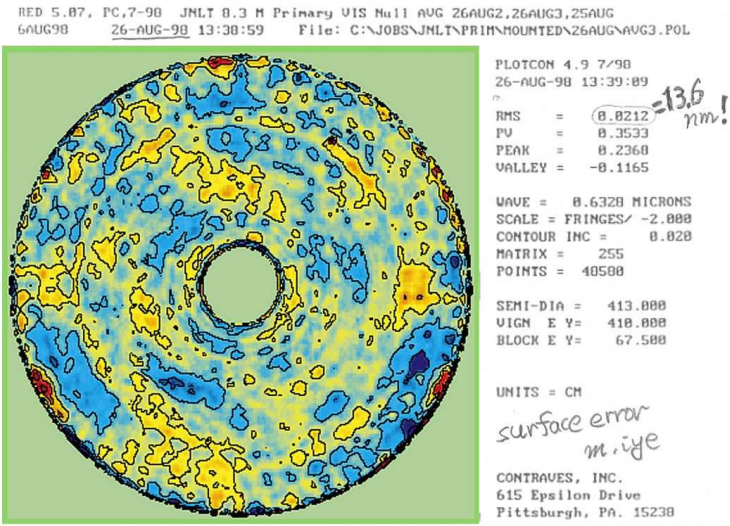}
\vspace{1mm} \caption{Surface figure error map of the finished primary mirror.  The residual figure error is as small as 14 nm rms. Three fixed points at the peripheral positions show larger residuals (NAOJ).}
\label{MirrorFinal}
\end{figure}

In 1994, the mirror blank was delivered to Contraves, near Pittsburgh, for polishing work. The mirror blank of 8.3 m in diameter and 30 cm thickness was then sagged down to a meniscus shape on a mold inside a heating furnace. It was then ground to a thickness of 20 cm. A total of 261 pockets of 15 cm depth were drilled from behind to house actuators. A careful acid-etching process was used to remove any microcracks. The final mirror had a 8.2 m diameter useful surface, which was polished to a hyperboloidal shape with a central curvature of radius of 30 m (a focal length of 15 m). Years of the careful polishing process ended with the final surface residual error, measured on the active support system in August 1998 at the acceptance test, as small as 14 nm rms after the active removal of low-order deformation components (Figure \ref{MirrorFinal}).

\subsection{\bf{Facilities}}
The enclosure of Subaru Telescope is located on the summit ridge of the Maunakea.  The coordinates of the telescope in the NAD83 system are at longitude ${W155^{\circ}28'34"}$ and latitude ${N19^{\circ}49'32"}$ with an altitude of 4163 m at the elevation axis. To prevent an uneven settlement of volcanic soil of the construction site, the soil cement method was used to improve the soil stiffness. The telescope pier of 14 m height from the ground to elevate the telescope above the ground layer was built on a round concrete mat slab of 43 m in diameter constructed 7 m underneath the ground.
The enclosure has its own foundation separated from of the telescope pier to avoid any vibrations induced from the enclosure.

A water tunnel test~\cite{Ando1991} and wind tunnel tests were used to compare the wind flow properties around various dome shapes.  A cylindrical shape design was chosen so as to avoid the turbulent ground-layer from flowing upward along the conventional semispherical dome. The Taisei Corporation designed and fabricated the enclosure to withstand an earthquake of 1000-year turnaround probability and a 70 m/s high wind.
The telescope structure was also designed to meet safety requirements.  Specially designed mechanical fuses were installed at the three fixed points of the primary mirror to release excessive force at these points if a strong earthquake was to occur. During construction, there was a fire accident and we regret that three local workers lost their lives.

The height of the telescope elevation axis at 24 m above the ground was chosen by taking the vertical distribution of turbulence due to the ground layer. 
The height of the enclosure is 43 m and the diameter of the enclosure rotation rail is 40 m. The total moving mass of the enclosure is 2,000 tons. The enclosure rotates in synchronization with the movement of the telescope. Various design considerations were made on thermal analysis and wind loading to ensure better image quality~\cite{Mikami1994}.

\begin{figure}[h]
\centering
\includegraphics[scale=0.5]{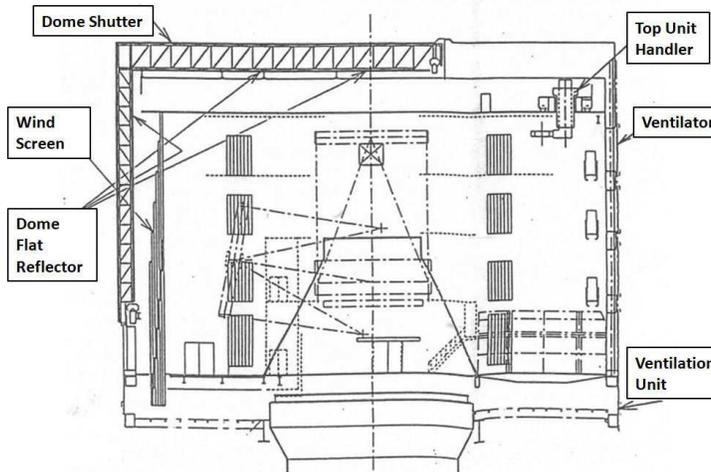}
\vspace{1mm} \caption{The Subaru Telescope enclosure has a biparting dome shutter and a wind screen. Several side windows are used to ensure the regulated wind passing inside the enclosure to flash internal turbulence and keep the inside temperature following the ambient temperature (NAOJ).}
\label{EnclosureSection}
\end{figure}

Two black painted ``Great Wall"s built on both sides of the telescope regulate the laminar wind flow through the telescope, and insulates the telescope thermally from other parts inside the enclosure.  The wind screen reduces high wind to the telescope; and several wind hatches help to regulate the wind flow, depending on the direction and speed of the night wind while observing. Figure \ref{EnclosureSection} shows a side view of the enclosure.

On one side, behind the great wall at the top unit floor, a rotary facility houses three secondary mirror units and two prime focus instruments. A specially designed top unit exchanger enables the safe exchange of these top units. Another facility device, a Cassegrain Instrument Automatic eXchanger (CIAX), helps to exchange Cassegrain instruments from the instrument-housing stations on the observing floor behind the great wall, where the cryogenic, electric, and network support keep the instruments for the standby condition ready for observations and maintenance~\cite{Usuda2000}.

The surface of the primary mirror is cleansed by a CO$_{2}$ dry-snow cleaning facility installed near the mirror cover every two weeks to keep the reflectivity at above 80\%. The primary mirror undergoes a re-aluminizing operation every three years. An 80-ton top crane of the enclosure lowers the mirror on its cell to the basement. A mirror-cleaning facility and an aluminizing chamber firing aluminum vapors on the mirror are used for the maintenance.

\subsection{\bf{Control and Software}}
The design of the Subaru Telescope control system consists of several segmented groups of computers connected to each other using UNIX-based Remote Procedure Call (RPC). The observational data are archived at the summit and transferred to the base facility at Hilo by an optical fiber link. The Subaru Telescope Archival System (STARS\footnote{ https://stars2.naoj.hawaii.edu/stars1min.html}) in Hilo is mirrored by a system at Astronomical Data Center of NAOJ in Mitaka. Any registered user can download desired observational data after the 18-month proprietary period using the SMOKA (Subaru-Mitaka-Okayama-Kiso Archive) system by a query server in Hawaii~\cite{Baba2002}. Subaru observational data are following the general rules of the Flexible Image Transport System (FITS) header standard.

\subsection{\bf{The First Light}}
The primary mirror was finished near Pittsburgh and shipped, via the Mississippi River and the Panama Canal to the Kawaihae port of Hawaii Island. In November 1998, it was transported to its final site using a special trailer vehicle equipped with computer-controlled jacks to avoid any excessive tilt of the load while ascending the mountain. The primary mirror was cleansed and aluminized to prepare for the first-light observation in the following month.

The engineering first-light was conducted on December 24, 1998. The first-light scientific observations made in January 1999 provided a superb image quality of the Subaru active optics telescope~\cite{Kaifu2000}. An inauguration ceremony was held on September 17, 1999, in the presence of Princess Norinomiya Sayako,  many international delegations, and the Representatives of the Native Hawaiian community on the summit of Maunakea.

\begin{figure}[h]
\centering
\includegraphics[scale=0.35]{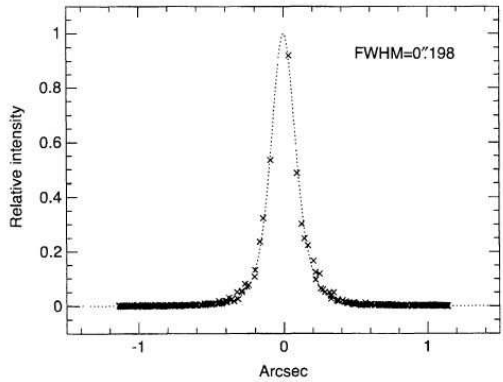}
\vspace{1mm} \caption{Point spread function of a star image taken by IRCS in the K-band under an excellent night condition shows a FWHM of 0.198 arcsec~\cite{Kaifu2000}. Note that the observation was made before the AO-system became available.}
\label{FWHM}
\end{figure}

The FWHM of the $K$-band image was confirmed to be as sharp as 0.198 arcsec (Figure\ref{FWHM}). The limiting magnitude in the $R$ band for a one-hour exposure by the Subaru Telescope reached as deep as 28-mag, comparable with that achieved by the Hubble Space Telescope (HST) from space for the equivalent band and exposure time~\cite{Iye2000}. The spatial resolution was twice better for HST due to the absence of atmospheric seeing, but the sensitivity was comparable because of the 13-times larger light-collecting power that contributed to enhancing the signal-to-noise ratio.

\vspace{5mm}
\section{\bf{Adaptive optics}}

\subsection{\bf{Early history}}
\begin{figure}[h]
\centering
\includegraphics[scale=0.4]{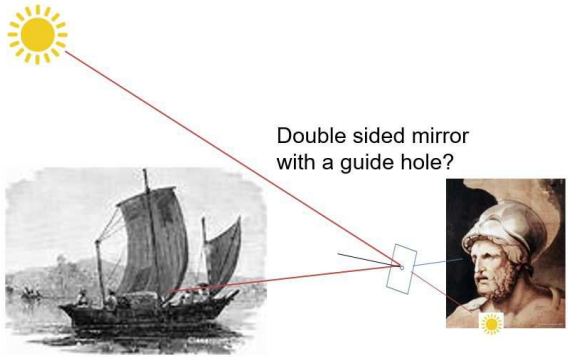}
\vspace{1mm} \caption{A possible manual alignment method to reflect the sunlight onto a war ship. A soldier can be trained to watch the sun spot produced through the hole on his cheek or cloth and guide the spot into the hole through the back reflection of the mirror.  This figure was reproduced based on the figure in ~\cite{Claus1973}.}
\label{Archimedes}
\end{figure}

Based on a Greek historian Lucian's description, Albert Claus suggested that Archimedes was the first to use adaptive optics by deploying Syracuse soldiers to line up and reflect sunlight with ``burning mirrors" to set Roman battleships on fire~\cite{Claus1973}. He conjectured that a handy size double-sided mirror with a central hole can be used to visually adjust its angle so that the specular reflection is pointing to the target (Figure \ref{Archimedes}). Though it is an interesting hypothesis, the availability of such mirrors at that time is highly dubious. The attack, if any, might have been made in a different way by firing flaming balls~\cite{Stavroudis1973}.

The California astronomer Hollace Babcock was actually the first to propose and develop a modern adaptive optics system concept to compensate for any atmospheric disturbance of the optical wavefront in real time to restore the diffraction-limited imaging capability~\cite{Babcock1953}.  The Russian meteorologist Valerian Tatarski mathematically expressed the characteristics of the wavefront perturbation generated while the light beam passes through a turbulent medium~\cite{Tatarski1961}.
D. L. Fried defined a scale length, called the Fried parameter, $r_{0}$, of a circular area within which the rms wavefront aberration due to a turbulent atmosphere is less than 1 radian~\cite{Fried1965}. The typical nighttime median, $r_{0}$, in meters at a wavelength of $\lambda$ ($\mu$m) for the zenith distance, $\beta$, can be approximated~\cite{Tyson2016} by

\begin{eqnarray}
[r_{0}]_{median} = 0.114 (\lambda/0.55)^{6/5} (\sec \beta)^{-3/5}   (m).
\end{eqnarray}

On a fair night at Maunakea, the Fried parameter for the 2.2 $\mu$m $K$-band is, therefore, about 0.4 m, while that for the 0.55 $\mu$m $V$-band is about 0.1 m.  Implementing adaptive optics in the $V$-band is much more difficult than in the $K$-band, since it requires 28-times more sensing and correcting elements and 5.3-times faster control for the same wind velocity.

It is important to note the similarity and difference of the active and adaptive optics~\cite{Beckers1993}.  The ``active optics" controls the wavefront distortions in a telescope introduced by mechanical, thermal, and optical effects in the telescope at frequency bandwidths slower than 1 Hz. Alternatively, the ``adaptive optics" controls rapidly varying atmospheric wavefront distortions, called ``astronomical seeing", at much higher band widths of 10--1000 Hz. ``Adaptive optics" should respond much more rapidly than ``active optics." Hence, the correction device should be much smaller and responsive than the primary mirror of the telescope, although the basic principle is common. This confusing terminology, however, is widely adopted due to historical reasons. 

Some developments were made for defense applications under strict information control during the 1970s and 1980s. The satellite tracking system of the US Air Force telescope on Haleakala was equipped with such an adaptive optics system. In 1990, soon after the launch of  HST, its primary mirror was unfortunately found to be mis-figured. An additional problem that NASA scientists spotted was unforeseen jitter vibration of the space telescope. The first generation of its solar panel boom was found to shrink/expand thermally every time the HST went in and out of the shadow of the Earth. The Air Force Maui Optical Site (AMOS) telescope with its adaptive optics was used to image the solar panel of the HST to measure the ``glint walking" along the bent boom from the specular reflection of the sunlight while the HST is staring at some observing target~\cite{Kissell1992}.

\begin{figure}[h]
\centering
\includegraphics[scale=0.4]{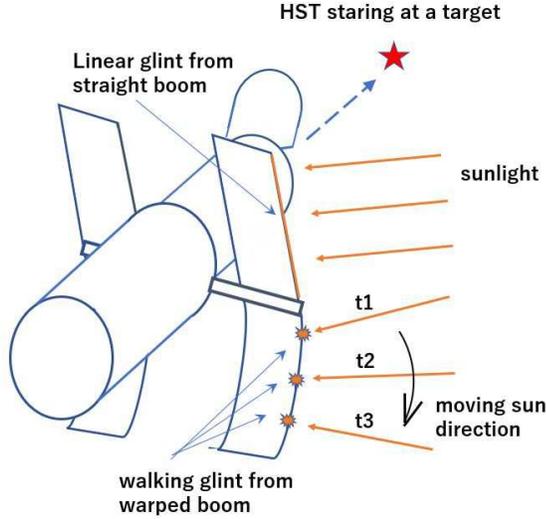}
\vspace{1mm} \caption{(Color online) The specular reflection of the sun light from HST, which is observing an observational target, was measured using adaptive optics imaging. The thermally warped HST solar array boom produces ``walking glints", in contrast to the line glint of the unbent boom. The figure was redrawn based on the original figure in ~\cite{Kissell1992}.}
\label{Kissell1992}
\end{figure}

Independent of the defense R\&D, the astronomical community also developed adaptive optics systems during the 1980s. ESO announced the first demonstration result where they used a 19-element piezo-driven deformable mirror coupled through a fast digital signal processor with wavefront sensing at 25 subapertures~\cite{Merkle1989}.

The wavefront error due to atmospheric turbulence can also be expressed by equation (1). By applying the modal correction from the lower-order modes by using adaptive optics, the residual wavefront error decreases from the uncompensated variance of $\sigma^{2}_{uncomp} = 1.03(D/r_{0})^{5/3}$ to tip-tilt the corrected variance of $\sigma^{2}_{tip-tilt} = 0.134(D/r_{0})^{5/3}$, where the number of corrected modes is $N_{m} = 2$. For a larger $N_{m}$, the corrected variance can be approximated by
 
\begin{eqnarray}
\sigma^{2} =0.2944N^{-\sqrt{3}/2}_{m}(D/r_{0})^{5/3}.
\end{eqnarray}

\begin{figure}[h]
\centering
\includegraphics[scale=0.4]{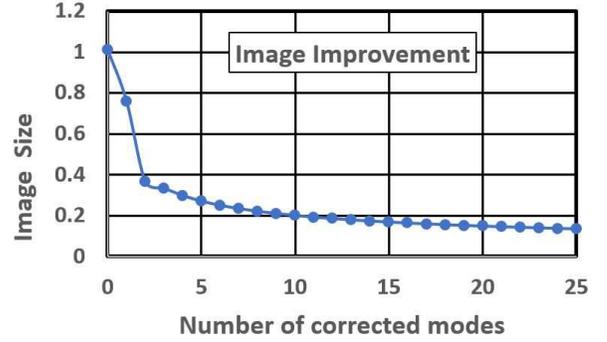}
\vspace{1mm} \caption{(Color online) The residual wavefront error, $\sigma$, due to turbulent air decreases as the number of corrected modes, $N_{m}$, increases.
$N_{m}$ = 2 corresponds to corrections of the tip and tilt only. $N_{m}$ = 3 corresponds to the tip-tilt plus defocus corrections. Next, corrections of astigmatism, coma, spherical aberration etc. follow as $N_{m}$ increases. With about 100 modes corrected, the image size sharpens to one-tenth of the size of the original blurred image.}
\label{CorrectionEffect}
\end{figure}

Figure \ref{CorrectionEffect} shows the image-sharpening achieved by increasing the number  $N_{m}$ of the controlled modes.

\begin{figure}[h]
\centering
\includegraphics[scale=0.48]{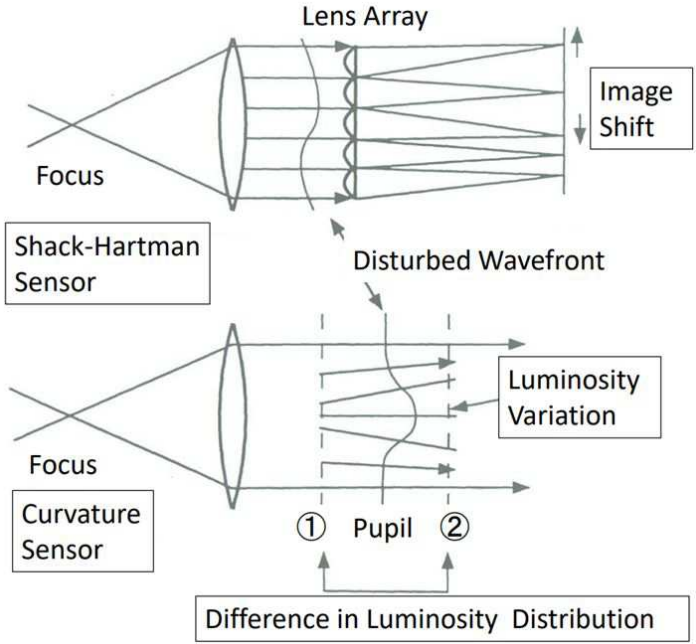}
\vspace{1mm} \caption{Shack-Hartman sensor and Curvature sensor. The Shack-Hartmann camera measures the displacement vectors of the individual subaperture images of the guide star, which refer to the local tilts of the wavefront. The curvature sensor measures the nonuniform illumination at the pupil plane that is produced by the local curvature of the wavefront. The actual measurement with a curvature sensor can be done by modulating the image plane of the sensor before (\MARU{1}) and after (\MARU{2}) the pupil plane. An additional vibrating mirror accomplishes this image plane shifting without moving the sensor. A provision to evaluate the difference between the two images is necessary.}
\label{WaveFrontSensing}
\end{figure}

Fran\c cois Roddier of University of Hawaii introduced a new idea concerning the curvature sensor to measure the wavefront~\cite{Roddier1988}. The Shack-Hartmann sensor measures the distribution of the first order derivative of the wavefront, namely, the local gradient vector, from the image displacements of a reference star for each subaperture.  In contrast, the curvature sensor measures the distribution of the second-order derivative of the wavefront, namely, the curvature distribution, from the illumination variations observed at the pupil plane. Figure \ref{WaveFrontSensing} shows the principles of the two methods.

A deformable mirror takes care of any wavefront correction. A thin membrane mirror, actuated by numerous rod-type piezoelectric actuators or by bimorph type piezoelectric actuators, is widely used. Excellent textbooks on adaptive optics are available~\cite{Roddier1999, Tyson2016}.

\subsection{\bf{Subaru Adaptive Optics System}}

Some early R\&D toward adaptive optics, including the development of an image stabilizer for the Okayama 188 cm telescope and an original stacked piezo-electric deformable mirror, were conducted at NAOJ and Communications Research Laboratory in the 1980s. Based on these experiences, the involved group developed the first-generation adaptive-optics system at the Cassegrain focus of the Subaru Telescope~\cite{Takami1996, Takami1999}. The strategy to implement second-generation adaptive optics at the Nasmyth focus of Subaru Telescope was established~\cite{Iye2004b}.

\begin{figure}[h]
\centering
\includegraphics[scale=0.4]{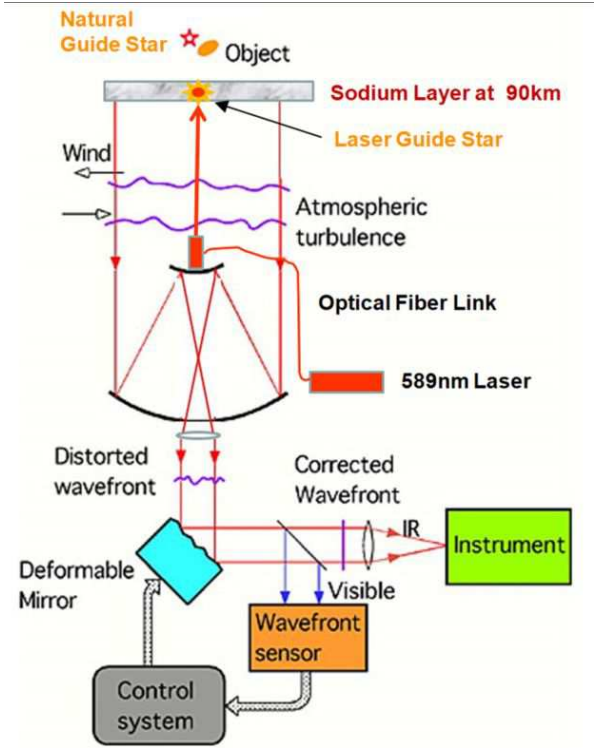}
\vspace{1mm} \caption{(Color online) The configuration showing the laser-guide-star adaptive optics system. The sodium laser beam produces a natural guide star shining at the sodium layer at 90 km height.  The wavefront sensor measures the optical disturbance at the kHz level, using either a natural star or a laser guide star. This information is used to drive the deformable mirror so as to compensate for any disturbance in real time (NAOJ).}
\label{AO-System}
\end{figure}

\begin{figure}[h]
\centering
\includegraphics[scale=0.5]{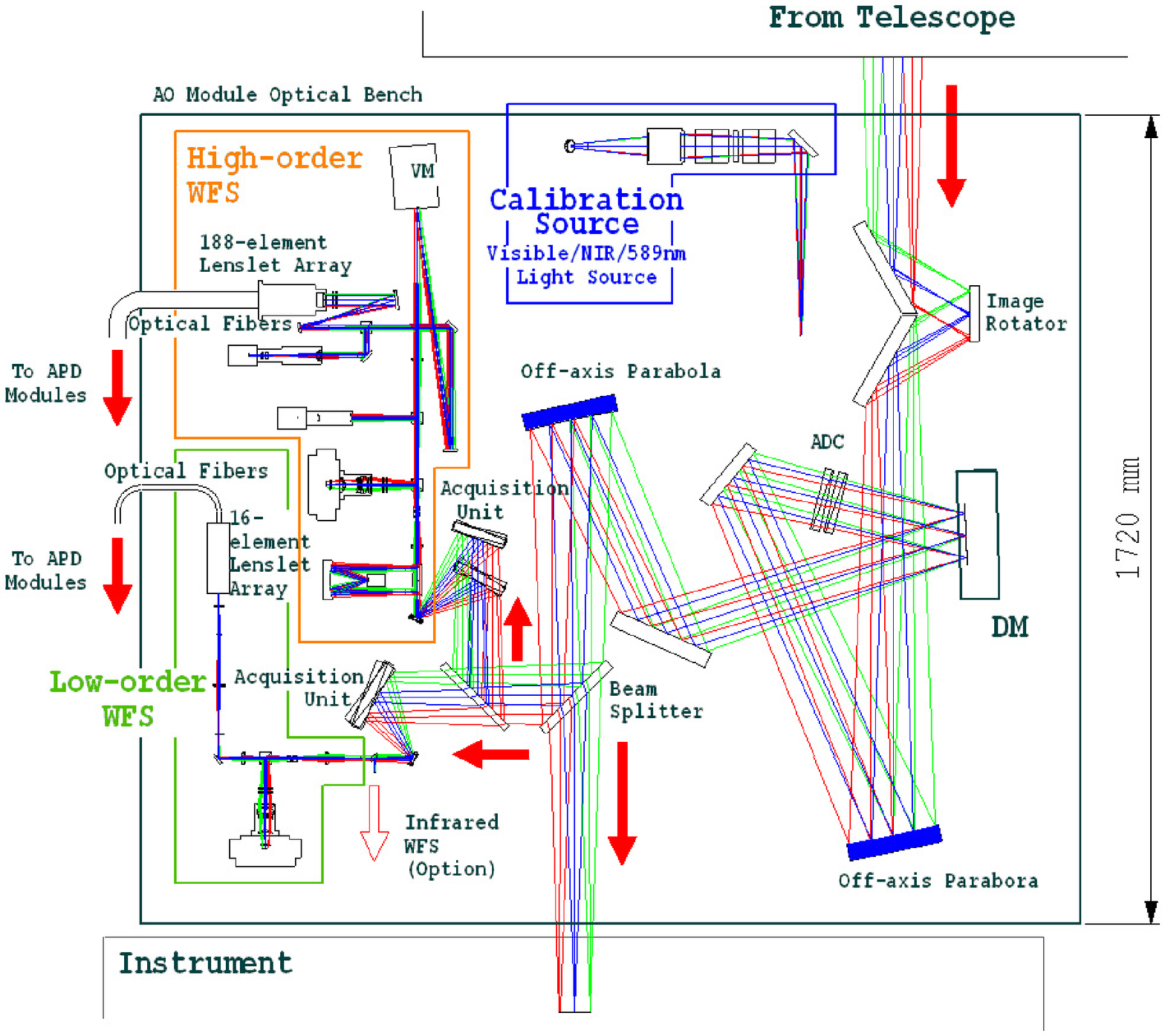}
\vspace{1mm} \caption{(Color online) Optical layout of the AO188 system. The incident beam from the telescope Nasmyth focus goes through the image rotator and the collimated beam is fed, through the ADC, to the deformable mirror for adaptive correcting of the wavefront error. The aberration-corrected beam is then delivered to the infrared scientific instrument. The optical beam selected by a beam splitter is fed to a high-order wavefront sensor and to a low-order wavefront sensor to derive the signal for driving the deformable mirror~\cite{Hayano2010}.}
\label{AO188Optics}
\end{figure}

MEXT's programs of Grant-in-Aids for Specially Promoted Research (2002--06) and for Scientific Research (S) (2007--11) enabled the implementation of the Laser Guide Star Adaptive Optics system with 188 control elements (LGSAO188) for Subaru Telescope.   Figure \ref{AO-System} shows a conceptual system outline of the LGSAO188. The actual layout of the AO188 optics is shown in Figure \ref{AO188Optics}~\cite{Watanabe2004}. The image rotator is a three-mirror optical device that de-rotates the object image rotation induced for the alt-azimuth telescope, like Subaru Telescope.  The atmospheric dispersion corrector (ADC) compensates the object image dispersed due to the chromatic dispersion of the atmosphere.  

A method to design the optimal configuration of the electrodes for a bimorph deformable mirror with 36 electrodes for the Cassegrain adaptive optics system was developed~\cite{Otsubo1996}. A new deformable mirror with 188 electrodes for the AO188 system of Subaru Telescope was fabricated by a French company, CILAS, based on the design by NAOJ~\cite{Oya2006}.

The real-time digital controller drives the bimorph mirror, using signals from the wavefront sensor, to compensate for any wavefront distortion at 1 kHz. The aberration-corrected beam is finally fed to the infrared Nasmyth instrument station.  The infrared instruments (HiCIAO, IRCS and others) can then make good use of the diffraction-limited image~\cite{Hayano2008}.  Olivier Guyon made an original addition of the extreme adaptive optics technology to enable higher level exoplanet detection~\cite{Guyon2018}.

\begin{figure}[h]
\centering
\includegraphics[scale=0.45]{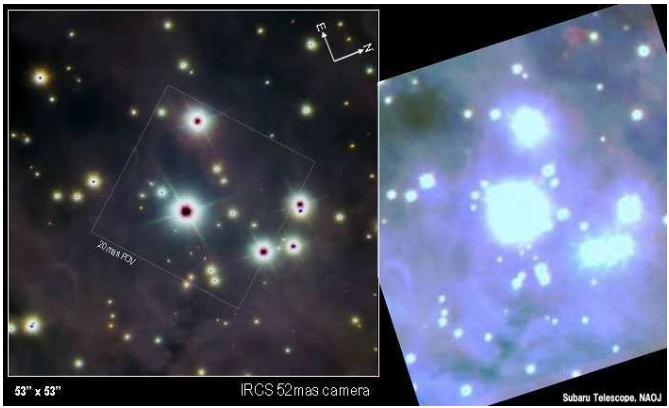}
\vspace{1mm} \caption{(Color online) Comparison of the Orion trapezium images with (left; taken in 2006) and without (right; taken in 1999) the adaptive optics system shows a significant image quality improvement by a factor of 10 (NAOJ).}
\label{trapezium}
\end{figure}

The first light of the Subaru AO188 system observation was made in 2006. The AO first-light image of the Trapezium in the Orion nebula revealed a sharp image size of 0".06 in the K-band, while a fair natural seeing image taken in 1999 was blurred 10 times to 0."6 (Figure \ref{trapezium}). A 7-million-dollar investment on the LGSAO188 improved the vision of the 400-million-dollar Subaru Telescope by an order of a magnitude.

\begin{figure}[h]
\centering
\includegraphics[scale=0.45]{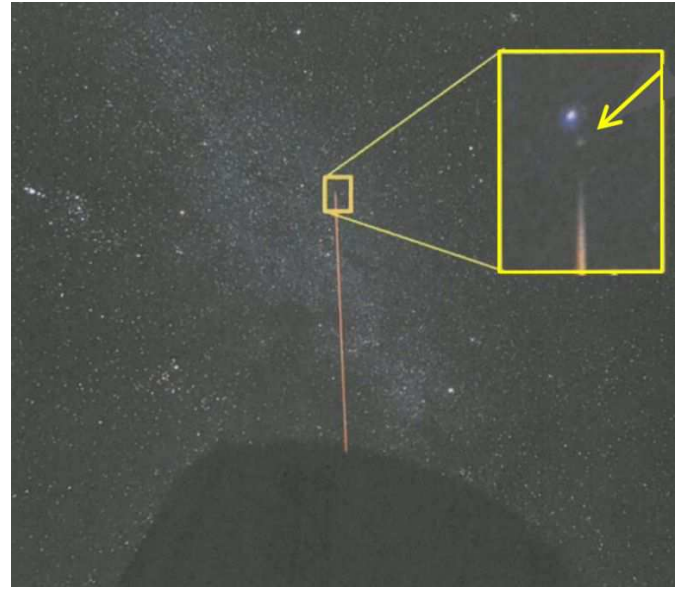}
\vspace{1mm} \caption{(Color online) Formation of a sodium laser guide star at 90 km in altitude was confirmed. Note the yellowish spot ahead of the diminishing point of Rayleigh scattering from the laser beam.} 
\label{LaserGuideStar}
\end{figure}

\subsection{\bf{Laser Guide Star}}

French astronomers Renaud Foy and Anton Labeyrie were the first in the astronomical community to mention the possibility of using the Rayleigh back-scattering light from a laser beam as a light source to probe turbulence in the atmosphere up to 100 km in altitude~\cite{Foy1985}. Laird Thompson and Chester Gardner of the University of Hawaii then reported their experiments at the Maunakea site to test the feasibility of using a laser to generate an artificial guide star in the mesospheric sodium layer~\cite{Thompson1987}.
The sodium layer is a layer of neutral atoms of sodium at 90 km above the sea level with a thickness of 5 km, produced by meteor ablation.  Sodium atoms excited by a 589 nm laser beam reemit the sodium D line emission.

Although not known to the astronomical community, similar research and experiments had been  conducted confidentially for defense applications. The ``detente" movement between the United States and the Soviet Union in the late 1970s and the fact that US tax was spent doubly for developing the same technology in astronomy and defense, enabled disclosing the classified technology, and subsequently papers on the laser-guide-star system were published~\cite{Fugate1991, Primmerman1991}.  

Yutaka Hayano developed a prototype sodium laser-guide-star system for Subaru Telescope~\cite{Hayano1999}, which evolved into a useful version in 2009~\cite{Hayano2008, Hayano2010}. Figure \ref{LaserGuideStar} shows the test generation of an artificial sodium laser guide star in 2006.

To observe using the adaptive optics system, a guide star bright enough ($R \le 11$) and close enough to the target (separation of less than 0.5 arcmin) is needed. Adaptive optics observations of exoplanets use a bright enough mother star as a natural guide star. Adaptive-optics observations of exoplanets use a sufficiently bright mother star as a natural guide star.  Adaptive-optics observations of extragalactic targets relied on a narrow chance of observing a natural guide star that is both bright and close enough to the target.  Such a chance was on average less than 1$\%$. 

The availability of a laser guide star increased the chance of making adaptive optics observations. LGSAO188 increased the sky coverage to about a $50\%$ level, but not to reach the $100\%$ level. The laser guide star generated from the telescope moves its position in the sky in a synchronous way to the vibrations of the telescope. Therefore, the jittery motion of the telescope cannot be measured using the laser guide star, itself. To measure the jittery motion of the telescope and make a relevant correction for it, monitoring a natural guide star, rather than a laser guide star, is indispensable. The required brightness for a natural guide star for tracking the telescope vibration is $R \le 19$, a much less stringent requirement for a guide star for full compensation of the atmospheric turbulence, $R \le 11$. Nevertheless, the chance of having a guide star close enough to the target of observation sets this limit on the availability of LGSAO188.  These limits can be relaxed by introducing higher power laser.

A transponder-based airplane detector was developed to power off the laser for the safety of any airplane flying at the nighttime across Maunakea.

\vspace{5mm}
\section{\bf{Science Instruments}}

The scientific instrument plan for the Subaru Telescope was discussed by the Subaru Advisory Committee in the early 1990s. Limited financial support for research and development studies was facilitated using the NAOJ fund. The Subaru Telescope and its instrumentation plan were first introduced at an international conference organized in Tokyo in October 1994~\cite{Iye1995}. 
A decision was made by the Subaru Advisory Committee in June 1995 to select seven first-generation instruments (Suprime-Cam~\cite{Miyazaki2002}, Faint Object Camera And Spectrograph (FOCAS)~\cite{Kashikawa2002}, High Dispersion Spectrograph (HDS)~\cite{Noguchi2002}, OH airglow Suppression spectrograph (OHS)~\cite{Iwamuro2001}, InfraRed Camera and Spectrograph (IRCS)~\cite{Tokunaga1998}, Coronagraphic Imager with Adaptive Optics (CIAO)~\cite{Tamura2000}, and COoled Mid-Infrared Camera and Spectrograph (COMICS)~\cite{Kataza2000}), and two observatory instruments (AO36~\cite{Takami1999}, MIRTOS~\cite{Tomono2000}) to be funded for construction among the proposed 14 instruments. 
The performances and actual usage of the initial set of seven open-use instruments were reported ~\cite{Iye2004a,Iye2004c}. 

Although the lifetime of a good telescope can be over five decades, the lifetime of scientific instruments is generally about a decade. This is because scientific needs are ever evolving, and the availability of new technology urges astronomers to build new instruments more suited for new scientific motivations.

\begin{table}[h]
\begin{center}
\vspace{1mm} \caption{Imaging capability of Subaru cameras. Column (4) refers to the maximum number of filters mountable on the holder units.}
\begin{tabular}{|c|c|c|c|} \hline
(1)&(2)&(3)&(4)\\
Instrument&FoV&Pixel Scale&Filters\\ \hline
HSC&$1^{\circ}.5\phi$&0".17&6\\
FOCAS&6'$\phi$&0".1&14\\
IRCS&21", 54"&0".020, 0".052&18\\
MOIRCS&4' x 7'&0".116&3\\
\hline
\end{tabular}

\label{table:Imagers}
\end{center}
\end{table}

\begin{table}[h]
\begin{center}
\vspace{1mm} \caption{Spectroscopic capability of Subaru spectrographs. Columns (3) and (4) refer to the slit width and the slit length, respectively.}
\begin{tabular}{lrrrl} \hline
(1)&(2)&(3)&(4)&(5)\\
Instrument&Resolution&Width&Length&Pixel\\ \hline
FOCAS&250&0".4&6'&0".1\\
&7500&0".4&multi slit&\\
HDS&32000&1"&10",60"&0".138\\
 &160000&0".2& &\\
IRCS&50&0".9&6.5-20"&0".020\\
 (grism)&14000&0".1&&0".052\\
IRCS&5000&0".54&3".5-9".4&0".055\\
 (echelle)&20000&0".14& &\\
MOIRCS&463&0".5&multi slit&0".116\\
&3020&0".5& &\\
\hline
\end{tabular}
\label{table:Spectrographs}
\end{center}
\end{table}

At the time of writing this review in 2021, the Subaru Telescope was serving five facility instruments (Hyper Suprime-Cam (HSC), FOCAS, HDS, IRCS, and Multi-Object Infra Red Camera and Spectrograph (MOIRCS)). A few other instruments are also offered by developer groups for open-use observers. Table \ref{table:Imagers} summarizes the imaging capability of these facility imaging instruments. Table \ref{table:Spectrographs} summarizes the spectroscopic capability of a facility instrument suite for 2021.

\begin{figure}[h]
\centering
\includegraphics[scale=0.4]{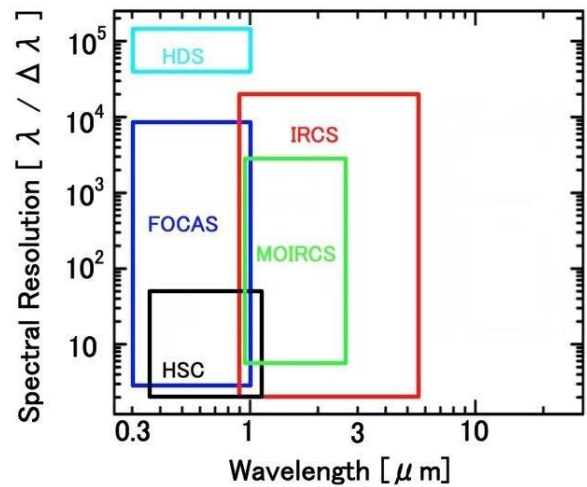}
\vspace{1mm} \caption{(Color online) Observational parameter space coverage of currently available open use instruments. The abscissa is the wavelength, and the ordinate is the spectral resolution NAOJ).}
\label{instrument-para}
\end{figure}

Figure \ref{instrument-para} shows the wavelength-spectral resolution parameter space covered by the currently available open-use instruments offered by the observatory.  
%Instead, SWIMS~\cite{Konishi2020}, a visiting instrument with similar capability is available on a risk-shared basis.

\subsection{\bf{HSC}}

Maki Sekiguchi made a pivotal development of mosaic CCD cameras~\cite{Sekiguchi1992} on the 105 cm Kiso Schmidt Telescope, the 4.2 m William Herschel Telescope and the 2.5 m Sloan Digital Sky Survey telescope~\cite{Gunn1998}. Satoshi Miyazaki established an evaluation system of CCD detectors and developed customized CCDs for astronomical applications under collaboration with Hamamatsu Photonics~\cite{Miyazaki1998}.

Miyazaki and his group built the Subaru Prime Focus Camera (Suprime-Cam) for integrating 10 devices of 4k$\times$2k CCDs to cover a 35'$\times$28' field of view~\cite{Miyazaki2002}. Its unique capability to cover a wide-field of view was a legacy not competed by any of the 8--10 m class telescopes.  Systematic projects that take advantage of the wide-field deep survey capability of the distant universe were successful. Some of the scientific achievements are discussed in the next section.

\begin{figure}[h]
\centering
\includegraphics[scale=0.3]{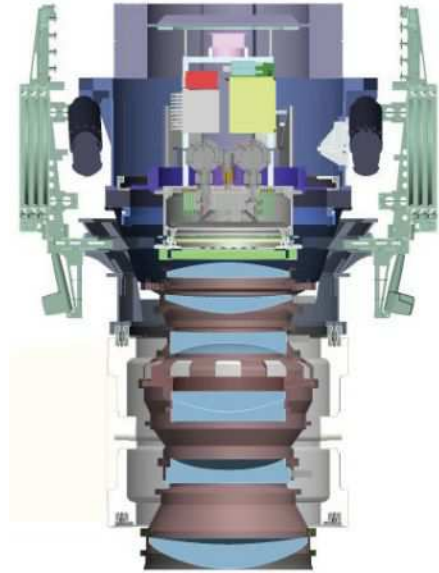}
\vspace{1mm} \caption{(Color online) Cross-section of the HSC~\cite{Miyazaki2012}. The wide-field corrector of HSC is shown in the lower part of the figure.}
\label{HSCSection}
\end{figure}

\begin{figure}[h]
\centering
\includegraphics[scale=0.4]{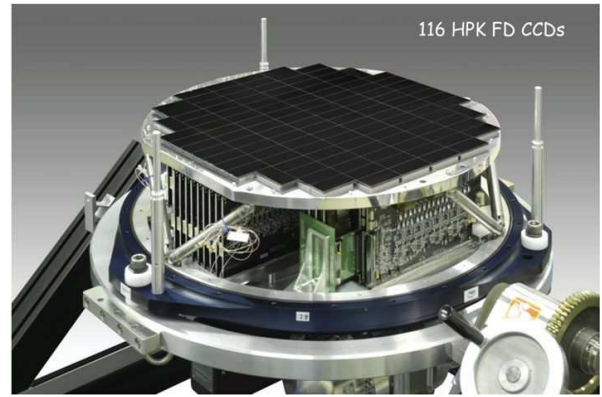}
\vspace{1mm} \caption{(Color online) 106 CCDs at the focal plane of HSC~\cite{Miyazaki2012}.}
\label{HSCCCDs}
\end{figure}

They built an even more ambitious camera, HSC~\cite{Miyazaki2018}, as a key instrument for the World Premier International Research Center Initiative. This project was jointly led by Kavli Institute for the Physics and Mathematics of the Universe (IPMU) and the NAOJ, including indispensable contributions from the High Energy Accelerator Research Organization (KEK), the Academia Sinica Institute for Astronomy and Astrophysics in Taiwan (ASIAA), and Princeton University. Figure \ref{HSCSection} shows the cross-section of HSC with its field corrector lens unit. The main optics was designed and fabricated by Canon. The wide-field camera has 104 devices of 4k$\times$2k CCDs deployed at the prime focus corrected for a $1.5^{\circ}$ field of view (Figure \ref{HSCCCDs}). The Wide-Field Corrector comprises seven lenses with five aspheric surfaces to deliver a high-quality image over the field for the wavelength region 420--1070 nm at F/2.23, with the function of atmospheric dispersion compensation. The filter exchanging unit of HSC accommodates six filters to enable imaging observation from a set of 5 broad-band filters and 18 narrow-band  filters. A total of 330 dark nights is dedicated for a systematic deep survey program (HSC-SSP; see section 7.2) using HSC to image millions of galaxies. The redshift of these galaxies is well inferred using a photometric redshift technique~\cite{Tanaka2018} to produce a 3D map of galaxies.
HSC is now the most heavily used instrument on the Subaru Telescope. 

\subsection{\bf{Faint Object Camera And Spectrograph}}
\begin{figure}[h]
\centering
\includegraphics[scale=0.48]{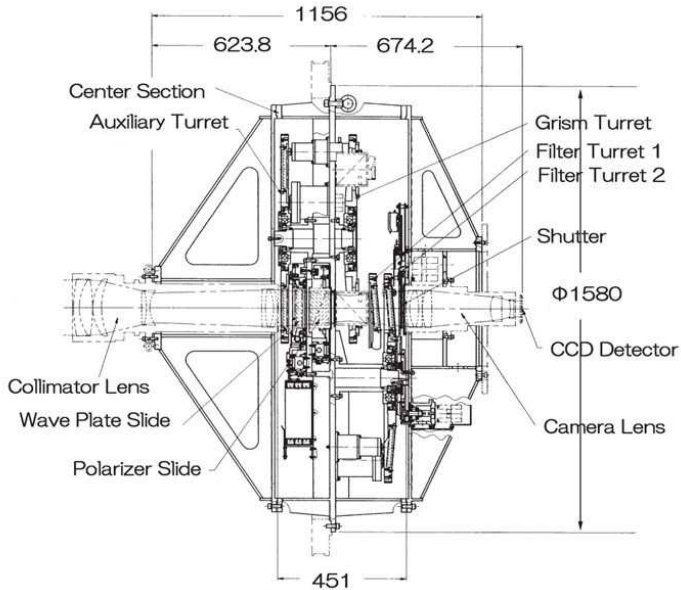}
\vspace{1mm} \caption{Faint Object Camera and Spectrograph (FOCAS) mounted beneath the Cassegrain focus~\cite{Kashikawa2002}}
\label{FOCAS}
\end{figure}

FOCAS~\cite{Kashikawa2002} enables the imaging of a 6' field of view with various filters, slit spectroscopy, and spectro-polarimetry for a wavelength region 370--1000 nm with a choice of grisms.  These grisms are direct vision dispersive elements designed by combining a transmissive dispersion grating and a prism~\cite{Ebizuka2011} so that the central wavelength of the dispersive band goes straight in the optical axis. The adoption of grisms as dispersive elements enables easy switching between imaging and spectroscopic observations by replacing filters and grisms (Figure \ref{FOCAS}). In addition to the long-slit spectroscopy, multi-object spectroscopy (MOS) for up to 40 objects in the field is possible. The MOS mask plate can be designed and prefabricated using the laser facility available in the summit enclosure.  The main optics and the body were manufactured by NIKON, and the CCD camera and the MOS unit were fabricated by NAOJ. 
%Redshift measurement of faint distant galaxies, spectral analysis of active galaxies, supernovae and gamma ray bursters are main science objectives of FOCAS.

\subsection{\bf{High Dispersion Spectrograph}}
\begin{figure}[h]
\centering
\includegraphics[scale=0.48]{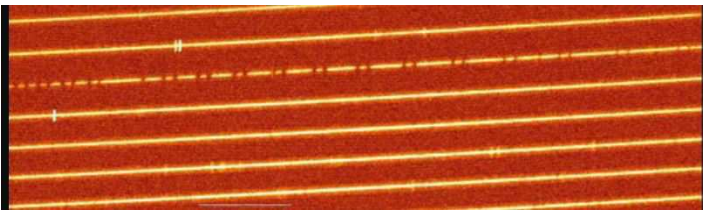}
\vspace{1mm} \caption{(Color online) Several orders of the echelle spectrum of a comet are shown.  The wavelength increases from left to right in each order and progressively upward in this image. The continuum spectrum shows the reflected sunlight from a comet with the absorption line. The emission lines from molecules, e.g., NH$_{2}$ etc., of the comet are superposed on the spectrum~\cite{Tajitsu2012}.}
\label{HDSComet}
\end{figure}

The HDS unit manufactured by NIKON ~\cite{Noguchi2002} is deployed at the optical Nasmyth platform for conducting high-resolution echelle spectroscopy at a wavelength range of 310--1000 nm.
Figure \ref{HDSComet} shows several orders of an echellogram image obtained by HDS. The highest spectral resolution of 160,000 is available for a slit width of 0".2, using an image slicer~\cite{Tajitsu2012}. HDS has been used for the abundance analysis of various types of stars, especially for metal-poor stars and for precise radial velocity measurements using an iodine cell for wavelength calibration~\cite{Sato2002}.

\subsection{\bf{InfraRed Camera and Spectrograph}}
\begin{figure}[h]
\centering
\includegraphics[scale=0.48]{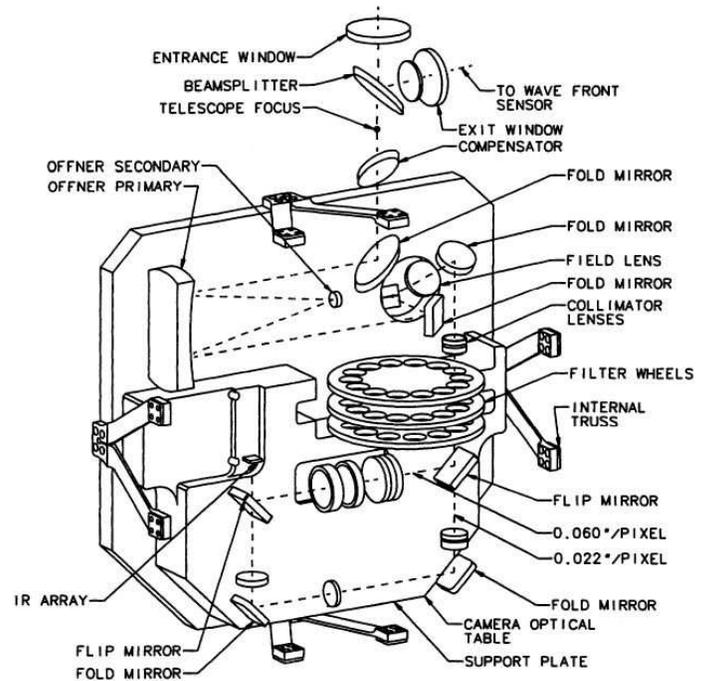}
\vspace{1mm} \caption{InfraRed Camera and Spectrograph (IRCS)~\cite{Tokunaga1998}.}
\label{IRCS}
\end{figure}

IRCS was designed and fabricated at the Institute for Astronomy, University of Hawaii, as an infrared Cassegrain instrument~\cite{Tokunaga1998, Kobayashi2000} (Figure \ref{IRCS}). IRCS offers infrared imaging and spectroscopic observation for  using an ALADDIN II 1024$\times$1024 InSb array. After completing the AO188 system, IRCS was transferred to the Infrared Nasmyth platform behind the AO188. Therefore, IRCS provides diffraction-limited imaging and spectroscopy at 20 mas/pixel and 52 mas/pixel sampling in the near-infrared wavelength 0.9--5.6 $\mu$m band with a choice of grism or an echelle disperser.

\subsection{\bf{Multi-Object InfraRed Camera and Spectrograph}}
\begin{figure}[h]
\centering
\includegraphics[scale=0.48]{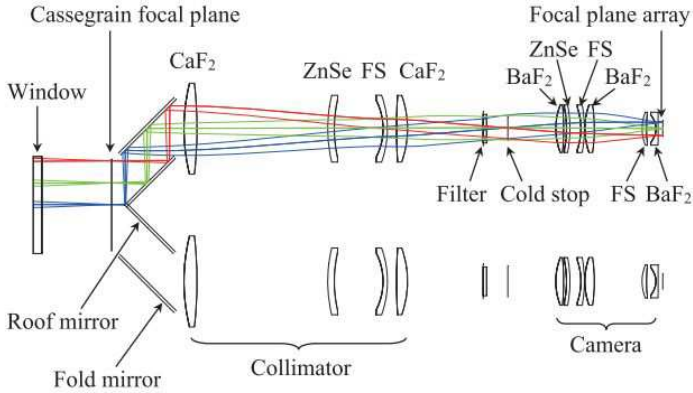}
\vspace{1mm} \caption{(Color online) Multi-Object InfraRed Camera and Spectrograph (MOIRCS)~\cite{Suzuki2008}.}
\label{MOIRCS}
\end{figure}

MOIRCS is a near-infrared instrument with the wide-field imaging mode and the MOS mode~\cite{Suzuki2008}. Dual optical trains cover a $4' \times 7'$ field of view for 0.85--2.5 $\mu m$ using two 2048$\times$2048 HgCdTe HAWAII-2 arrays (Figure \ref{MOIRCS}).

\subsection{\bf{Other Instruments}}

So far, ten early generation instruments of Subaru Telescope have been decommissioned. The list includes AO36, CIAO, COMICS, HiCIAO, CISCO/OHS, FMOS, Kyoto3DII, RAVEN, and Suprime-Cam. Some of these were replaced with new generation instruments. 

Eight visiting instruments are under operation for Subaru Telescope. Six of these (CHARIS~\cite{Currie2018}, FastPDI, IRD~\cite{Kotani2020}, MEC, REACH~\cite{Kotani2020}, and Subaru Coronagraphic Extreme Adaptive Optics (SCExAO)~\cite{Jovanovic2015}) are dedicated instruments using Subaru AO188.   SWIMS~\cite{Konishi2020} is a near-infrared multi-object spectrograph developed for the 6 m telescope of the University of Tokyo Atacama Observatory. The design and development of these new instruments are carried out by various groups in and outside Japan with close arrangements using the Subaru Telescope. Information on these instruments is available on the Subaru Web Page\footnote{https://subarutelescope.org/Observing/Instruments/index.html}.

%\subsection{\bf{Data Archive}}

\vspace{5mm}

\section{\bf{Major Scientific Achievements}}
During 2000--20, 2282 refereed papers were published around Subaru Telescope, of which 51\% were headed with Japanese first authors and 78\% include international coauthors. The distribution of the number of papers by journal is ApJ(41\%), MNRAS(17\%), PASJ(16\%), AJ(8.0\%), A\&A(7.8\%), ApJS(3.4\%), PASP(1.2\%), Icarus(0.8\%), Nature(0.8\%), Science(0.6\%), and others(3.8\%).  

The distribution of the number of papers by the instrument is Suprime-Cam(37.6\%), HSC(11.6\%), HDS(11.4\%), FOCAS(10.5\%), IRCS(9.6\%), CIAO/HiCIAO(6.1\%), MOIRCS(5.4\%), COMICS(3.7\%), CISCO(2.7\%), FMOS(2.4\%) and OHS(1.0\%).
Suprime-Cam and HSC produced just half of the refereed papers. 

The educational impact was also outstanding, producing 156 PhDs. The total citation counts on the refereed papers amount to 110,841. The subjects involving science cover from the solar system to the deep universe. Described below are some examples of selected topics. 

\subsection{\bf{Epoch of Cosmic Reionization}}

According to the major current model, the Universe started with inflation and a big bang 13.8 billion years ago, which is followed by a rapid expansion and cooling.  At 0.38 million years after the Big Bang, at a redshift of $z\sim1100$, the temperature of the Universe had dropped to about 3000 $K$, cooled enough for protons and electrons to combine to form neutral hydrogen atoms. From this epoch, the photons were decoupled from further cooling matter. Since there were not yet any celestial shining bodies, this cold early Universe is referred to as the ``dark age." The current scenario of structure formation expects gravitational growth of dark-matter fluctuation to exist, leading to the formation of primordial galaxies at around $z\sim20$. UV radiation from the newly formed massive stars and quasars would have warmed up the intergalactic hydrogen, and ``cosmic reionization" would have occurred.

SDSS spectroscopic observations have shown from ``Gunn-Peterson" tests of rest-frame UV spectra of high-redshift quasars that the reionization process was completed by $z =6$~\cite{Fan2006}. Observational studies for $z\ge6$ became practical with the advent of Subaru Telescope.

\paragraph{\bf{Subaru Deep Field}}
Nobunari Kashikawa and his group launched a large program to make a deep galaxy survey over a blank field, named Subaru Deep Field (SDF), using Suprime-Cam imaging in five broad-band filters ($B, V, R, i', z'$-bands) and two narrow-band filters (NB816 and NB921~\cite{Kashikawa2004}). The objective was to construct a deep galaxy catalog containing more than $10^5$ galaxies with photometry and follow-up spectroscopy for the selected subsamples. Two narrow-band filters were designed to pick up photometrically the Lyman $\alpha$ emitters at redshifts of $z = 5.7$ and $z = 6.6$ for which their Lyman $\alpha$ emission would fall in the narrow-band paths of NB816 and NB921. The results of Lyman $\alpha$ emitter survey observations were reported for $z = 5.7$~\cite{Shimasaku2006} and for $z = 6.6$~\cite{Kashikawa2006}, respectively. 

\begin{figure}[h]
\centering
\includegraphics[scale=0.55]{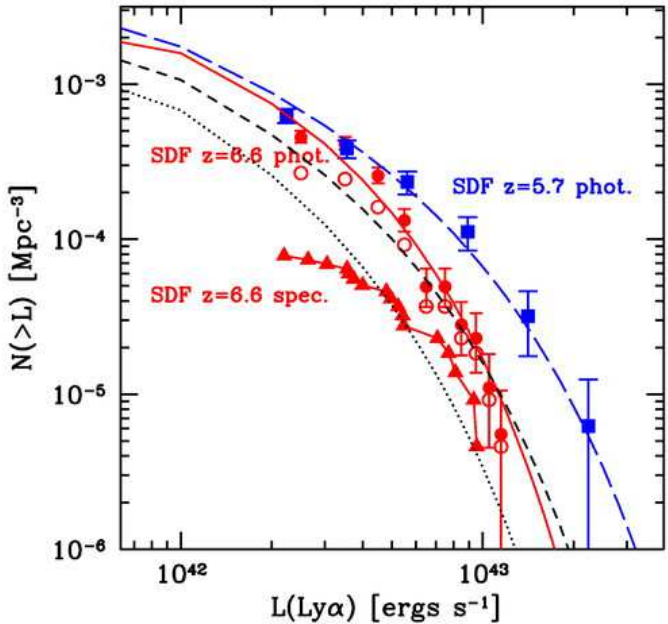}
\vspace{1mm}
 \caption{(Color online) Cumulative Ly$\alpha$ luminosity functions, as observed in the SDF. The squares represent the photometric luminosity function for $z = 5.7$, and the long-dashed curve shows its fitted Schechter function. Filled circled show the completeness corrected photometric luminosity function for $z=6.6$, and the solid curve shows its fitted Schechter function. The triangles are the spectroscopic luminosity function for $z = 6.6$~\cite{Kashikawa2006}.}
\label{KashikLF}
\end{figure}

Kashikawa highlighted the significant decrease in the luminosity function of the Lyman $\alpha$ emitter population from $z = 5.7$ to $z = 6.6$ (Figure \ref{KashikLF})~\cite{Kashikawa2006}. They ascribed it to the end of reionization of the intergalactic medium. 

\begin{figure}[h]
\centering
\includegraphics[scale=0.4]{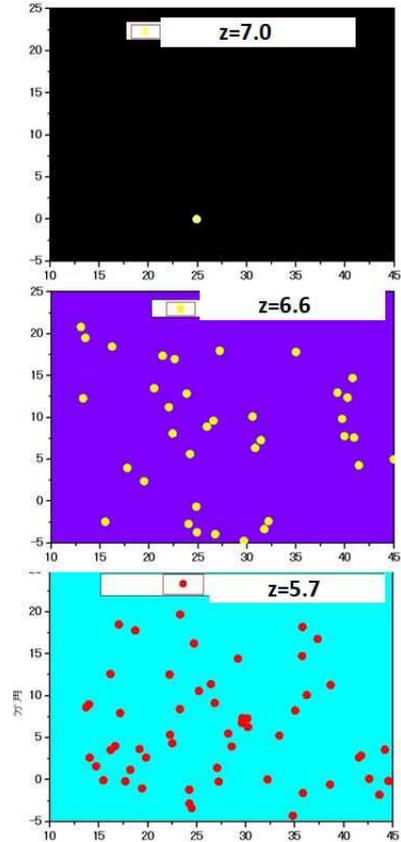}
\vspace{1mm} \caption{(Color online) The distribution of Lyman alpha emitters, as confirmed in the SDF at three epochs, $z$ = 5.7~\cite{Shimasaku2006}, 6.6~\cite{Kashikawa2006}, and 7.0~\cite{Iye2006}, with limiting narrow-band magnitudes of 26.0, 26.0 and 24.9, respectively. Although the sensitivity for detection is 1.1-magnitude more shallow for $z$-7.0, the clear drop in the number of Lyman alpha emitters indicates cosmic reionization proceeding significantly at $6.6 \le z \le 7.0$.}
\label{reionization}
\end{figure}

The author of this paper and his group made an additional survey at $z = 7.0$ using NB973 and reported finding a single source, IOK-1, at $z = 6.964$, that remained the most distant galaxy spectroscopically confirmed until 2011~\cite{Iye2006}.
Figure \ref{reionization} compares the distribution of Lyman $\alpha$ emitters at three epochs:  $z = $5.7,  6.6, and 7.0.  An abrupt drop in the number density between redshifts 6.6 and 7.0 as is shown in Figure\ref{reionization} indicates that the transparency of the intergalactic medium against Lyman $\alpha$ photons changed during this period.

\paragraph{\bf{Tension on the Reionization Epoch}}
Masami Ouchi derived a luminosity function, clustering measurements, and Ly$\alpha$ line profiles based on 207 Ly$\alpha$ emitters at $z$ = 6.6 to conclude that major reionization occurred at $z_{re} \ge 7$~\cite{Ouchi2010}.

An early analysis of the cosmic microwave background data obtained by Wilkinson Microwave Anisotropy Probe (WMAP) reported the derivation of cosmological parameters, that showed the epoch of cosmic reionization at $z_{re} = 10.5 \pm1.2$~\cite{Larson2011}, somewhat discrepant with the results obtained from Ly$\alpha$ emitters census~\cite{Iye2011}. However, a recent update based on Planck 2018 results show $z_{re} = 7.64\pm0.74$~\cite{Aghanim2020}, consistent with the Ly$\alpha$ census result. The early discrepancy between the two methods is thus resolved.

%\begin{figure}[h]
%\centering
%\includegraphics[scale=0.5]{GN-z11.eps}
%\vspace{1mm} \caption{(Color online) Confirmation of the carbon emission lines [CIII]1907 and  CIII]1909 at a redshift z=11 for GN-z11~\cite{Jiang2020}.}
%\label{GN-z11}
%\end{figure}

\paragraph{\bf{The Most Distant Galaxies}}
The Ly$\alpha$ emitter IOK-1 at $z = 6.964$~\cite{Iye2006} remained the highest redshift galaxy during 2006.9--2010.12. In 2010, all of the top 10 entries of the most distant galaxies spectroscopically confirmed were monopolized by Subaru Telescope discoveries. This was the high time that Subaru Telescope proved its capability in the wide-field survey.
After IOK-1 lost the top position, GN-108036 at $z = 7.213$~\cite{Ono2012} and XSDF NB1006-2 at $z = 7.215$~\cite{Shibuya2012} retrieved the world record position during 2012.1--2012.6 and 2012.6--2013.10, respectively.

Since the Ly$\alpha$ emission from sources beyond $z = 7.3$ woud enter the infrared wavelength region where the Si CCD detector loses sensitivity, objects at higher redshifts could only be studied by infrared instruments for which the field of view is much narrower than for Suprime-Cam or HSC.  The infrared camera mounted on HST has been used to identify high-redshift candidates from infrared photometry.
Some of those objects were verified spectroscopically to be at high redshifts. 

Inoue \etal\space \rm  made the first successful detection of the [OIII] emission line of SXDF NB1006-2 using the submillimeter interferometer ALMA~\cite{Inoue2016}.  This approach produced further discoveries of MACS 1149-JD1 at $z = 9.1096$~\cite{Hashimoto2018} and MACS J0416.1-2403 at $z = 8.3118$~\cite{Tamura2019}, respectively. 
At the time of writing this review, GN-z11 is the furthest galaxy spectroscopically confirmed at $z$=10.957 based on its emission lines, [CIII]$\lambda$1907 and CIII]$\lambda$1909~\cite{Jiang2020}.

%At the time of writing this article, Table \ref{table:top10} shows the top ten list of most distant galaxies spectroscopically confirmed.

%\begin{table}
%\begin{center}
%  \vspace{1mm} \caption{Top 10 list of most distant galaxies}
%    \begin{tabular}{rlrl}
%\hline
%\#&Object &$z$&Paper \\
%\hline\hline
%1 &GN-z11&10.957&~\citet{Jiang2020} \\
%2 &MACS1149-JD1&9.110&~\citet{Hashimoto2018} \\
%3 &EGSY8p7&8.683&~\citet{Zitrin2015} \\
%4 &A2744$\_$YD4&8.382&~\citet{Laporte2017} \\
%5 &MACS0416$\_$Y1&8.312&~\citet{Tamura2019} \\
%6 &EGS-zs8-1&7.730&~\citet{Oesch2015} \\
%7 &z7$\_$GSD$\_$3811&7.664&~\citet{Song2016} \\
%8 &z8$\_$GND$\_$5296&7.508&~\citet{Finkelstein2013} \\
%9 &EGS-zs8-2&7.477&~\citet{Stark2017} \\
%10 &SXDF-NB1006-2&7.215&~\citet{Shibuya2012} \\
%\hline
% \end{tabular}\\
% \label{table:top10}
%\end{center}
%\end{table}

\subsection{\bf{Dark Matter and Dark Energy}}
Supernova cosmology~\cite{Perlmutter1999} and cosmic microwave background observation~\cite{Larson2011} established the presence of dark energy and dark matter. Subaru Telescope contributed supernova cosmology nobel prize with FOCAS spectroscopy, including the one at $z$ = 1.35, the most distant supernova spectroscopically confirmed with ground-based telescopes at that time~\cite{Morokuma2010}.

Toward elucidating the actual distribution of dark matter, the wide-field, high spatial-resolution, multi-band imaging of HSC is most relevant for the gravitational weak-lensing analysis of the dark matter distribution at a high redshift universe. 

The Hyper Suprime-Cam Subaru Strategic Program (HSC-SSP) was designed to make 300 nights of observation at over 1400 deg$^{2}$ field in five broad bands ($grizy$) down to a $5\sigma$ limiting magnitude of $r\sim26$~\cite{Aihara2018a}.  An additional 30 nights were assigned to this program in 2019. The main goals are to study the distribution of galaxies, the evolution of the large-scale structure, studies on dark matter and dark energy, and the local galactic archeology. The first data release for the first 1.7 years of observations for 61.5 nights yielded a catalog containing a total of 10 million objects for UltraDeep, Deep, and Wide fields~\cite{Aihara2018b}. For galaxies, the five-band photometry of individual objects enabled a photometric redshift estimation by comparing the observed spectral energy distribution with those derived from stellar evolution models of various ages and redshifts.

\begin{figure}[h] 
\centering
\includegraphics[scale=0.4]{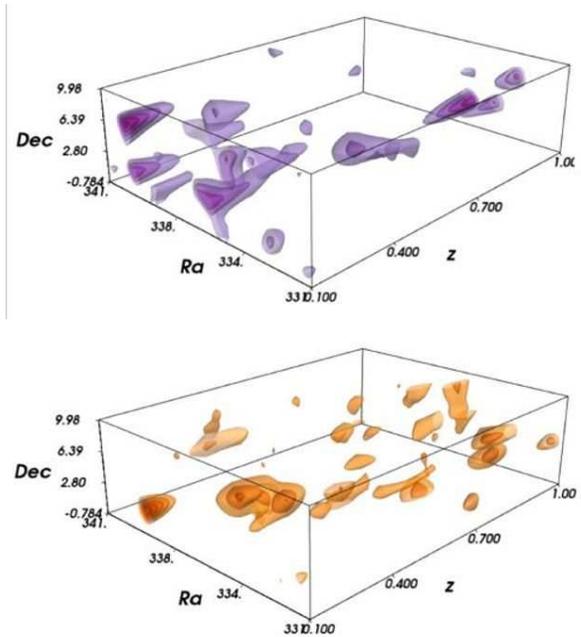}
\vspace{1mm} \caption{(Color online) (upper) Three-dimensional dark matter mass map derived from weak-lensing analysis. (lower) The corresponding three-dimensional galaxy mass map derived from the photometric luminous red galaxies sample. The enhanced regions with contours from 2$\sigma$ to 6$\sigma$ are shown~\cite{Oguri2018}. Note that condensation of dark matter in the universe grows from high redshift to low redshift.}
\label{DMdistribution}
\end{figure}

\paragraph{\bf{Weak-Lens Mapping of Dark Matter}}
The weak-lensing effects of distant galaxy shapes from the first-year HSC imaging data (167 deg$^{2}$) were measured to reconstruct the mass distribution over a wide solid angle~\cite{Oguri2018}. 
Figure \ref{DMdistribution}, obtained by combining the photometric redshift information, suggests the redshift evolution of dark matter aggregation on a cosmological scale (Figure \ref{DMdistribution}). 

\begin{figure}[h]
\centering
\includegraphics[scale=0.35]{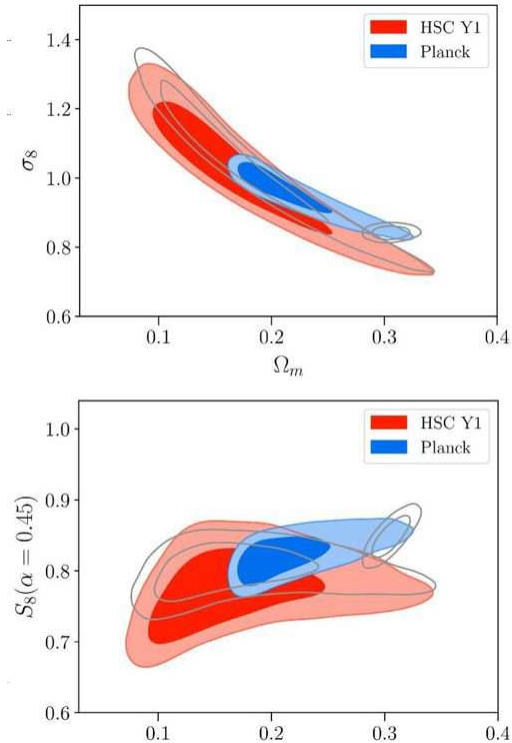}
\vspace{1mm} \caption{(Color online) Constraints of cosmological parameters in the $\Omega_{m}$-$\sigma_{8}$ plane (upper) and in the $\Omega_{m}$-$S_{8} (\alpha = 0.45)$ plane (lower) for $\omega$CDM model. Refer to the paper~\cite{Hikage2019} for details.}
\label{cosmology-tension}
\end{figure}

\paragraph{\bf{Tension on Cosmological Parameters}}
Despite the great success of the standard $\Lambda$ cold dark matter ($\Lambda$CDM) model to describe the expansion history and the growth of the large-scale structure of the Universe, the exact nature of the dark matter and dark energy remains unknown. The weak gravitational lensing effect provides a measure to compare the observed mass distribution of the Universe with those predicted from various models to identify most plausible cosmological parameters. The analysis of the power spectrum of the weak-lensing effect, in four tomographic redshift bins (z = 0.3--1.5) of 9 million galaxies with shape parameters and photometric redshifts derived from the 90 nights of HSC observation, was carried out~\cite{Hikage2019}. A careful comparison with various mock models of $\Lambda$CDM led to an independent derivation of the cosmological parameters $\Omega_{m}$ and $S_{8}$ from the local Universe ($z\sim1$), which are delicately discrepant from those values derived from cosmological microwave background (CMB) analysis ($z\sim1000$) (Figure \ref{cosmology-tension}).

Similarly, a measurement of the Hubble constant, using a time-delay analysis of gravitationally lensed quasars by HST and ground-based telescopes with adaptive optics was performed~\cite{Wong2020}. They found $H_{0} = 73.3^{+1.7}_{-1.8}$km s$^{-1}$ Mpc$^{-1}$, in agreement with local measurements of $H_{0}$ from type Ia supernovae, but in $3.1\sigma$ tension with the value $H_{0} = 67.4\pm0.5$km s$^{-1}$ Mpc$^{-1}$ derived from Planck CMB observations.

Although various approaches to derive cosmological parameters have been converging to give much more consistent results during these two decades, some tensions still appear between the results obtained from a $z\sim1000$ universe and from a $z\sim1$ universe.

\paragraph{\bf{Primordial Black Holes}}
The existence of a significant amount of dark matter is a common understanding. However, apart from its non-baryonic and non-relativistic nature, the identification of dark matter remains one of the most important quests in modern physics.  Experimental searches for weakly interacting massive particles (WIMPs) have not been successful. One of the alternative candidates, theoretically proposed, is primordial black holes (PBH) that formed in the early universe.  If a large number of such PBHs exist in the universe, they should be detected through the microlensing effect they would produce when they pass in front of stars. 

\begin{figure}
\centering
\includegraphics[scale=0.45]{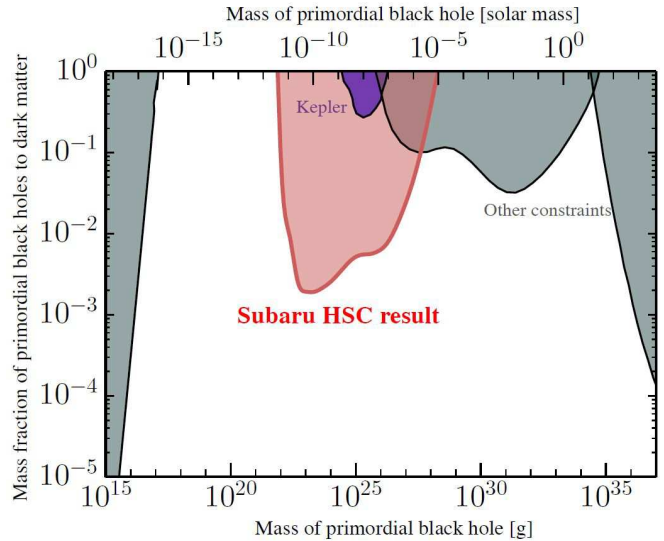}
\vspace{1mm}
\caption{(Color online) New stringent limits on the mass spectrum of low-mass primordial black holes derived from observations of microlensing events of stars in M31~\cite{Niikura2019}. The figure was redrawn by M.Takada.}
\label{PBH}
\end{figure}

A unique observation of a 7-hour monitoring experiment of a hundred million stars in the neighboring galaxy M31 with HSC was made to search for microlensing events due to low-mass PBHs~\cite{Niikura2019}. They found only one candidate event, compared to the expected 1000 events if PBHs make up all dark matter in the MW and M31 halo regions. As a result, for the first time, they obtained stringent upper limits on the abundance of PBHs in the mass range of $[10^{-11}, 10^{-6}] M_\odot$ (Figure \ref{PBH}).

\subsection{\bf{Gravitational Wave Events, Quasars, Supernovae}}

\begin{figure}[h]
\centering
\includegraphics[scale=0.5]{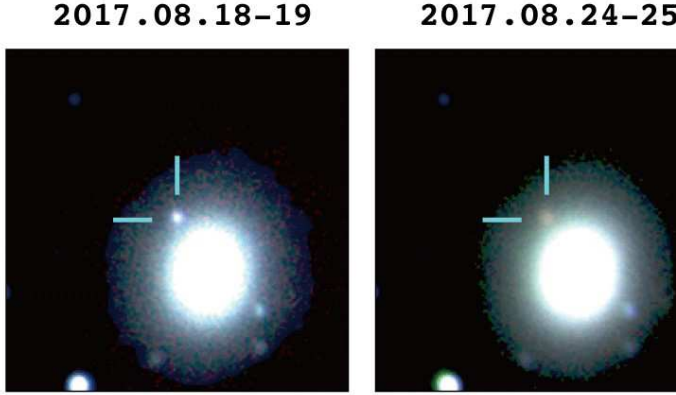}
\vspace{1mm} \caption{(Color online) Optical counterpart of the gravitational wave event GW170817 observed two days and eight days after the explosion. These pseud-color images were produced from $z$-band image taken by the Subaru/HSC and $H$ and $K_{s}$ band images taken by SIRIUS/IRSF~\cite{Utsumi2017}.}
\label{GW170817}
\end{figure}

\paragraph{\bf{Binary Neutron Star Merger}}
The detection of gravitational waves from merging binary black holes was one of the great scientific discoveries during the 2010s. A real breakthrough was another detection of a gravitational wave event, GW170817 from merging binary neutron stars~\cite{Abbott2017}. Identifying an optical counterpart of the gravitational wave event GW170817 was an epoch-making discovery, which was also contributed by Subaru observations~\cite{Utsumi2017, Tominaga2018}. Figure \ref{GW170817} shows two images of GW170817 taken two days and eight days after the event, showing the decay in luminosity and change in color. 

A theoretical study showed that optical and near-infrared light curves of the optical counterpart of GW170817, lasting for more than 10 days, are explained by $0.03 M_{\odot}$ of ejecta containing lanthanide elements~\cite{Tanaka2017}. Their radiative transfer simulations of kilonova, where the radioactive decays of $r$-process nuclei synthesized in the merger power optical and infrared emissions, are compatible with the observations. This result indicates that neutron-star mergers are the origin of the $r$-process elements in the Universe. 

\begin{figure}[h]
\centering
\includegraphics[scale=0.48]{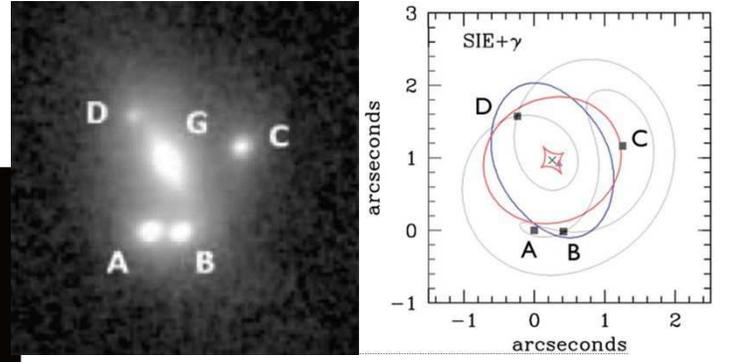}
\vspace{1mm} \caption{(Color online) (Left) LGSAO188 image of SDSS J1330+1810 with four lensed images, A-D, and a lensing galaxy, G. (Right) Gravitational lensing model of SDSS J1330+1810. The cross shows the position of the lensing galaxy, and the squares show the positions of lensed images of a quasar~\cite{Rusu2016}.}
\label{Lensedqso}
\end{figure}

\paragraph{\bf{Lensed Quasar}}
LGSAO analysis of 28 lensed quasars elucidated their configurations of the lensing galaxies and the lensed images~\cite{Rusu2016}. Among them, SDSS J1310+1810 is a quasar at a source redshift of $z_{s} = 1.393$, which is strongly lensed by the gravitational field of a lensing galaxy at $z_{l} = 0.373$ in front of the quasar, forming four clear images within a 1.8 arcsec separation. With precise astrometric and photometric data on four images, one can constrain the gravitational-lensing models, as well asthe mass and shape of the lensing galaxy. Figure \ref{Lensedqso} shows the observed image and the constructed lensing model of SDSS J1330+1810. A close analysis revealed a need to add satellite galaxies in addition to the main lensing galaxy in order to precisely reproduce the observed results. Monitoring these objects to measure the time delays among the lensed images can be an independent method to derive the Hubble constant.

\begin{figure}[h]
\centering
\includegraphics[scale=0.35]{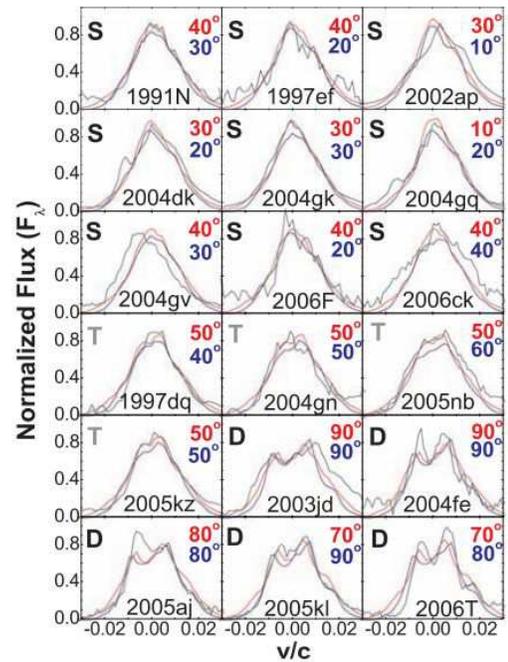}
\vspace{1mm} \caption{(Color online) Observed [OI]$\lambda$6300, 6363 emission-line profiles (black curves), classified into characteristic profiles: single-peaked (denoted by S, top nine SNe), transition (T, middle four SNe), and double-peaked (D, bottom five SNe). Line profiles of the bipolar model with different viewing directions are shown for a highly aspheric model (red curves with the direction denoted by the red text) and for a relatively less aspheric model (denoted in blue curves and text)~\cite{Maeda2008}.}
\label{OI-Profiles}
\end{figure}

\paragraph{\bf{Asymmetric Explosion of Supernovae}}
Stars more massive than 10 solar mass end their lives when the nuclear fuel in the innermost region is burnt out. Without sufficient pressure to balance the gravity, their core collapses to a neutron star or a black hole.  FOCAS spectroscopy of some core-collapse supernovae (CC-SNe) show a highly aspherical explosion, and suggest their link to long-duration gamma-ray bursts (GRBs)~\cite{Maeda2008}. Many of the observed oxygen emission lines show double-peaked profiles, a distinct signature of an aspherical explosion (Figure \ref{OI-Profiles}). They conclude that all CC-SNe from stripped-envelope stars are aspherical explosions, and that SNe accompanied by GRB exhibit the highest degree of aspheric nature.

%\begin{figure}[h]
%\centering
%\includegraphics[scale=0.44]{SNIa-spec.eps}
%\vspace{1mm} \caption{(Color online) (Left) Comparison of light curves for SN2002bo showing high velocity gradient (HVG: $\ge$ 70km $s^{-1} d^{-1}$) and SN1998bu showing low velocity gradient (LVG: $\le 70km s^{-1} d^{-1}$)
%. Peak luminosity is 1.8 mag brighter and the magnitude drop is faster for SN2002bo. (Right) Absorption line profiles of Si II 635.5 nm 4days before and 28 days after the $B$-band maximum. Figures are taken from~\cite{Maeda2010}.}
%\label{SNIa-spec}
%\end{figure} 

%Type Ia supernovae are known to have uniform light curves with peak luminosity well calibrated by the magnitude decline rate. SN Ia are, hence, used in supernova cosmology as standard candles to measure the distance. However, recent investigations revealed that the true nature of type Ia supernovae is more complicated.~\citet{Maeda2010} found the variety in their spectroscopic behavior can be understood as the manifestation of off-center explosion of SN Ia viewed from different angles.

\subsection{\bf{Galaxy interactions}}
\begin{figure}[h]
\centering
\includegraphics[scale=0.48]{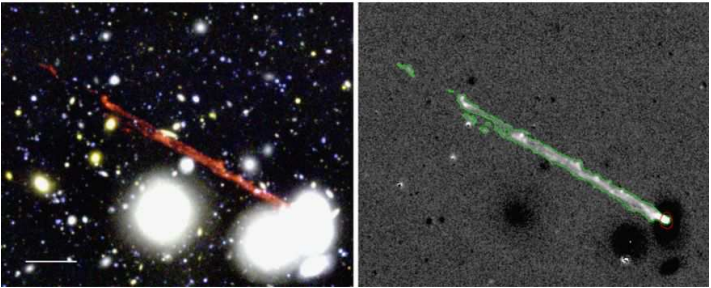}
\vspace{1mm} \caption{(Color online) (left)$B$,$R$ and $H\alpha$ color composite. (right) $H\alpha-R$ image showing $H\alpha$ emission extending out to 60 kpc from the parent galaxy GMP2910 in the Coma Cluster~\cite{Yagi2010}.}
\label{yagijet}
\end{figure}

\paragraph{\bf{Gas streak from galaxies}}
Multi band imaging observations revealed an unusual complex of narrow blue filaments, bright blue knots, and H$\alpha$ emitting filaments extending up to 80 kpc from an E+A galaxy RB199 in the Coma Cluster~\cite{Yoshida2008}. Also reported is a finding of H$\alpha$ filaments from 14 galaxies in the Coma Cluster~\cite{Yagi2010}. Figure \ref{yagijet} shows an impressive collimated H$\alpha$ filament extending 60 kpc from the core of the galaxy GMP2910. The origin of such a structure is hitherto enigmatic, but is likely to be related to the ram-pressure stripping of gas clouds from a high-speed collision of a galaxy with a hot intracluster medium.

\paragraph{\bf{Stellar Population of Dwarf Galaxies}}
Dwarf galaxies, found during the last decade or two in the Local Group, are relevant fossil objects to study the formation and evolutionary history of galaxies. 

Detailed studies concerning the stellar population of ultra-faint dwarf galaxies became feasible by comparing the observed color-magnitude diagrams with various stellar population models with different ages and star-formation history~\cite{Okamoto2012}. The high spatial resolution needed to resolve individual stars and the high sensitivity of the Subaru Telescope enable studying the color-magnitude diagram well below the main-sequence turn-off point (MSTO).

\begin{figure}[h]
\centering
\includegraphics[scale=0.48]{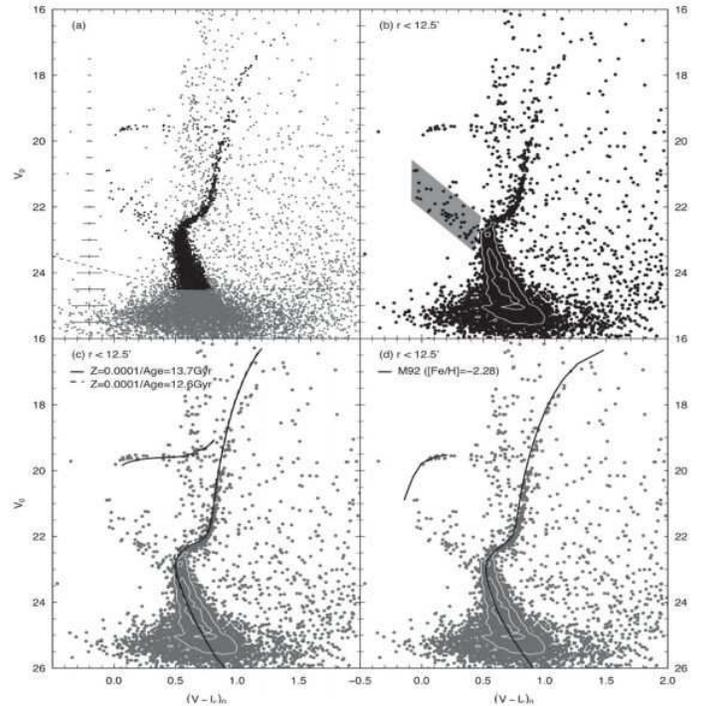}
\vspace{1mm} \caption{Color-magnitude diagrams of a Bo\"{o}tes ultra-faint dwarf spheroidal galaxy. (a) Entire region, (b) Central region, (c) Model tracks with different ages fitted, and (d) Model track for M92 overplotted~\cite{Okamoto2012}.}
\label{Okamoto2012}
\end{figure}

Okamoto \etal\space \rm found, for instance, Bo\"{o}tes I ultra-faint dwarf  spheroidal galaxy has no intrinsic color spread in the width of its MSTO, suggesting that it is composed of a single old population (Figure \ref{Okamoto2012}). This finding indicates that the gas in their progenitor was depleted more efficiently than those of brighter galaxies during the initial star-formation phase. The spatial distribution of the stellar population indicates a tidal interaction with our Milky Way Galaxy.

\begin{figure}[h]
\centering
\includegraphics[scale=0.5]{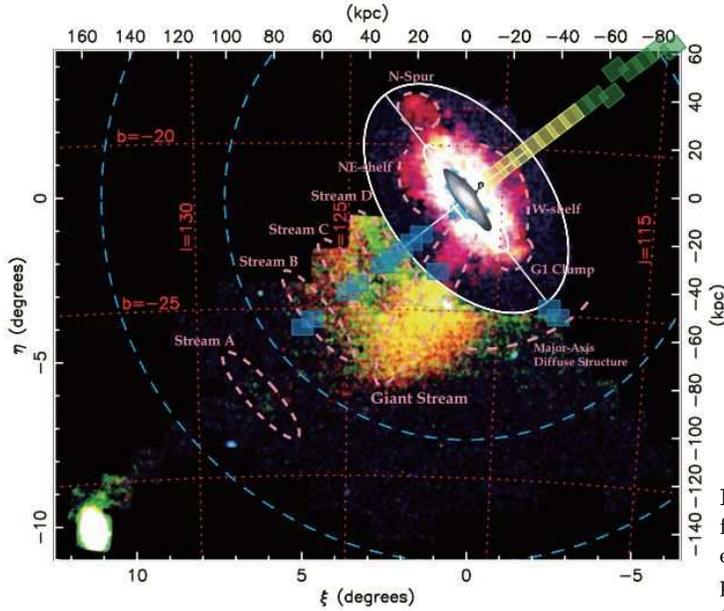}
\vspace{1mm} \caption{(Color online) Locations of surveyed fields (rectangular fields) around M31 overlaid on a color-coded stellar density map (taken from Figure 50 of the Ibata et al. 2007). Red, green, and blue colors show, respectively, stars with -0.7 $<[Fe/H]<$0.0, -1.7$<[Fe/H]<$-0.7, and -3.0$<[Fe/H]<$-1.7. The locations of the Giant stream and a few other streams are shown~\cite{Tanaka2010}.}
\label{M31Halo}
\end{figure}

\paragraph{\bf{M31 outer halo}}
An extensive imaging survey of the stellar halo population of M31 was made along its minor axis out to 100 kpc (Figure \ref{M31Halo}), and two new substructures were found~\cite{Tanaka2010} in the northwest halo in addition to the two substructures reported in the southeast halo~\cite{Ibata2007}.

These substructures are probably disrupted remains of dwarf galaxies trapped into M31 in the past. The color magnitude diagrams of the southern Giant Stream show high metallicity peaked at [Fe/H]$\ge$-0.5 and an age of $\sim$8 Gyr. The metallicity of those substructures is not uniform, indicating that the halo population has not had sufficient time to dynamically homogenize the accreted populations. The faint and smooth metal-poor stellar halo extends up to the survey limit of 100 kpc. To probe the outer halo distribution of matter and dark matter, further observation with higher sensitivity is needed.

\paragraph{\bf{Milky Way halo and dwarf galaxies}}
%The big bang nucleosynthesis (BBN) predicts the production of Li in the first several minutes and Li abundance has been studied to check the BBN process.
%~\citet{Aoki2009} derived, with HDS spectroscopy, Li abundances for eleven metal-poor turn-off stars and found the average Li abundance for stars with [Fe/H] $\le$ -3 is lower by 0.3 dex than that of stars with higher metallicity. This finding implies the existence of some unknown process to deplete Li in metal-poor stars.

\begin{figure}[h]
\centering
\includegraphics[scale=0.35]{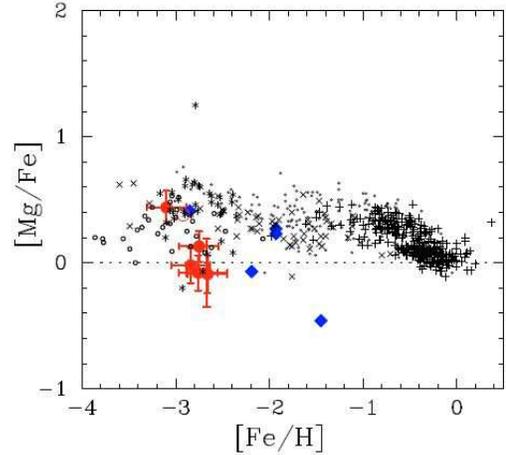}
\vspace{1mm} \caption{(Color online) Abundance ratios [Mg/Fe] as a function of the iron abundance [Fe/H]. Red filled circles with error bars are new measurements for six extremely metal-poor Sextans stars. Measurements for Galactic halo stars are shown by open circles, asterisks, crosses, and small circles, while plus symbols are for disk stars~\cite{Aoki2009A}.}
\label{Sextans}
\end{figure}

HDS spectroscopy of six extremely metal-poor red giants in the Sextans dwarf spheroidal galaxy was conducted~\cite{Aoki2009A}. They found that their [Mg/Fe] abundance ratios at the low metallicity end ([Fe/H] $\le$-2.6) are significantly different from those of the main part of the Galactic halo (Figure \ref{Sextans}). These results imply that classical dwarf spheroidal galaxies, like Sextans, do not resemble the building blocks of the main part of the Galactic halo. Further studies are needed on the roles of dwarf galaxies in the early Galaxy formation.

\begin{figure}[h]
\centering
\includegraphics[scale=0.35]{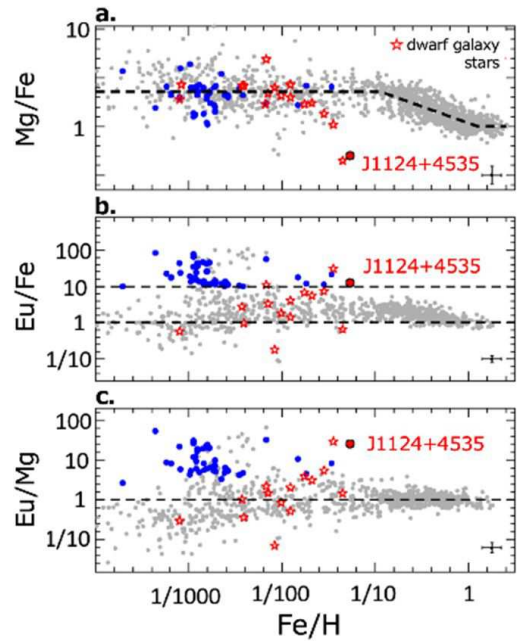}
\vspace{1mm} \caption{(Color online) Abundance ratios [Mg/Fe] and [Eu/Mg] as shown in panels a and c, respectively, of J1124+4535 (red circle) appear consistent with those measured for stars in Ursa Minor dwarf galaxy (red open star symbols)~\cite{Xing2019}. Figure was redrawn by W. Aoki.}
\label{abundance}
\end{figure}

Another HDS spectroscopy of a metal-poor star J1124+4535, with 1/20 solar metallicity, discovered by LAMOST observation, was performed~\cite{Xing2019}. They found that its abundance ratio of the r-process element, Eu, with respect to Fe is more than one order of magnitude larger than that of the Sun, while the $\alpha$ elements show a striking deficiency. Atomic abundance ratios (Figure \ref{abundance}) like this star are found in stars of the Ursa Minor dwarf galaxy, which suggests that this star was originally a member star of a dwarf galaxy that was merged in our Milky Way Galaxy.

\subsection{\bf{Exoplanets}}
\begin{figure}[h]
\centering
\includegraphics[scale=0.48]{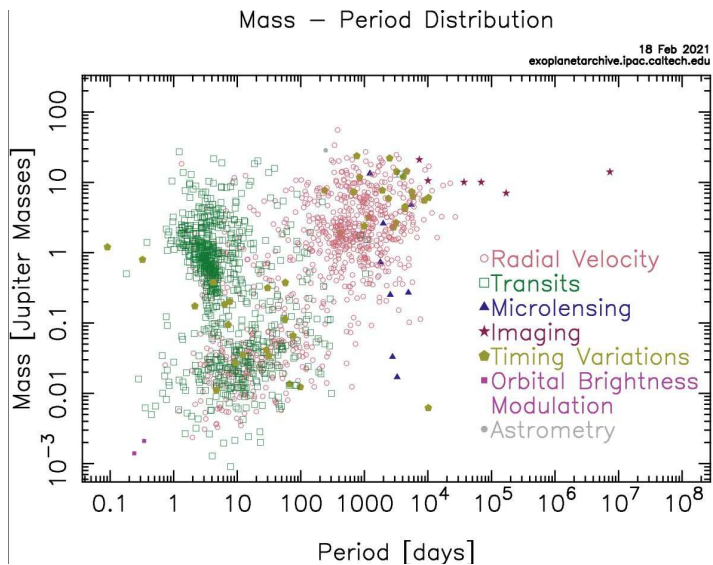}
\vspace{1mm} \caption{(Color online) Distribution of exoplanets in the orbital period vs mass plane, showing their variety and detection biases, depending on observational methods. This figure was reproduced from NASA Exoplanet Archive.}
\label{exoplanets}
\end{figure}

The discovery of a Jupiter-mass planet by measuring the radial-velocity modulation of 51Peg, caused by the orbital motion of the planet, opened a new look into exoplanet studies~\cite{Mayor1995}. The Kepler mission~\cite{Borucki2010}, finding exoplanets by monitoring their transits in front of mother stars, increased the number of detected exoplanets by well over 1,000 within 16 months of operation~\cite{Batalha2013}. The pulsar timing method~\cite{Wolszczan1992} and the microlensing method~\cite{Beaulieu2006} are additional methods used to detect exoplanets.  With a rapid increase in entries in the list, it is now a common understanding that many stars have planets. The NASA exoplanets archive now registers 4352 exoplanets. Figure \ref{exoplanets} shows the period-mass distribution of detected exoplanets. Chushiro Hayashi\footnote{Member of the Japan Academy in 1987--2010} established a legacy scenario (Kyoto Model) to explain the physical processes to form the Solar system~\cite{Hayashi1985}. The discovery of various planet systems, including hot Jupiters that are not expected in the Kyoto Model, highlights the need for more versatile planet formation schemes.

\paragraph{\bf{Imaging Exoplanets}}
Motohide Tamura reviewed the SEEDS project, direct imaging program to study exoplanets and proto-planetary disks, using HiCIAO and AO188 of the Subaru Telescope~\cite{Tamura2016}. To date, 12 undisputed exoplanets have been imaged with the help of adaptive optics systems, including some from the Subaru SEEDS program.

The first discovery of a brown dwarf or a planet with Subaru/HiCIAO imaging observation was brought for an object at a projected separation of 29 au from a Sun-like star GJ758~\cite{Thalmann2009}. The $H$-band luminosity of GJ758B suggests a temperature of  549--637 $K$ and a mass of 12--46 $M_{J}$, depending on its assumed age of 0.7--8.8 Gyr.

\begin{figure}[h]
\centering
\includegraphics[scale=0.45]{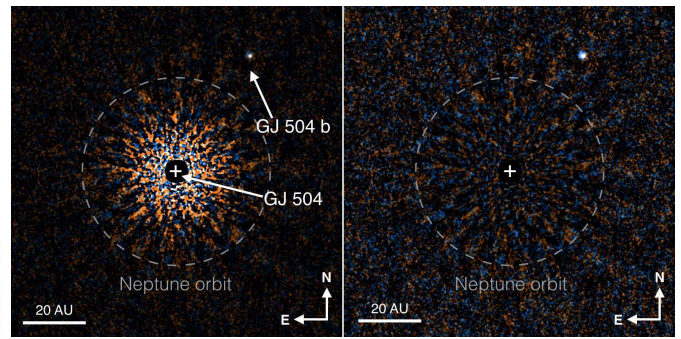}
\vspace{1mm} \caption{(Color online) Exoplanet GJ504b (upper right). The HiCIAO coronagraph optics masks the bright central star, GJ504, so as to enhance the dynamic range of vision for fainter objects~\cite{Kuzuhara2013}.}
\label{GJ504b}
\end{figure}

The discovery of a Jovian planet around a Sun-like star, GJ504 (Figure \ref{GJ504b}), followed. They derived an age of $160^{+350}_{-60}$ Myr, a mass of $4^{+4.5}_{-1.0} M_{J}$, and a temperature $510^{+30}_{-20} K$ for this planet~\cite{Kuzuhara2013}. There remains, however, some dispute about its system age.

\begin{figure}[h]
\centering
\includegraphics[scale=0.45]{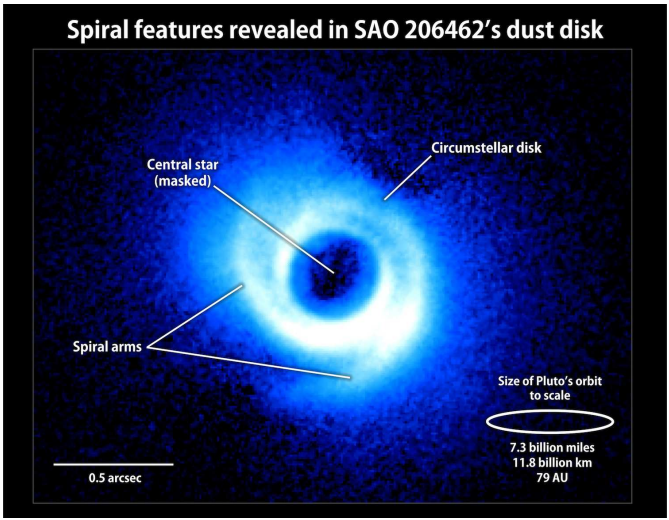}
\vspace{1mm} \caption{(Color online) Spiral structure in the circumstellar disk around a Herbig Ae star SAO206462 revealed by HiCIAO~\cite{Muto2012}.}
\label{SAO206462}
\end{figure}

The discovery of a spiral structure in circumstellar disks of AB Aur and SAO 206462(HD135344B) (Figure \ref{SAO206462}), respectively~\cite{Fukagawa2004, Hashimoto2011, Muto2012} opened a new challenge concerning the formation mechanism of such a structure. A similar spiral structure and ring structure with gaps are found in other young stars with HiCIAO, and now with ALMA. The formation mechanisms of these spiral and ring structures are being discussed in terms of the density wave theory.

To enable even a higher resolution and contrast, a team led by Olivier Guyon  is developing a new camera, SCExAO~\cite{Jovanovic2015}. The SCExAO is a high-contrast high-efficiency coronagraph exploiting Phase Induced Amplitude Apodization (PIAA) Coronagraph. It is placed behind the AO188 and in front of various new scientific instruments to enable the imaging of exoplanets close to their mother star.

COMICS 10 $\mu$m long-slit spectroscopic observation of the edge-on proto-planetary disk around $\beta$ Pic~\cite{Okamoto2004} brought a finding of substructure. They fitted the observed spectra with a model composed of three emission components from: $0.1 \mu$m glassy amorphous silicate grains, 2.0 $\mu$m glassy olivine amorphous grains and Mg-pure crystalline ollivine (forsterite, Mg$_{2}$SiO$_{4}$) grains. The distribution of crystalline grains and micron-size grains were continuous. They, however, found that the 9.7 $\mu$m emission from $0.1 \mu$m amorphous silicate grain shows three peaks indicative of the presence of planetesimal ring belts.

\paragraph{\bf{Dynamics of Exoplanet Systems}}
Hirano \etal\space reported a joint analysis of Subaru spectroscopy to measure the Rossiter-McLaughlin (RM) effect and the Kepler photometry during the planet transits of the Kepler Object of interest (KOI)-94 system~\cite{Hirano2012}. The RM effect is a spectroscopic shift of the stellar radial velocity observed due to a partial eclipse of the stellar surface by the transit of the planet, which conveys information concerning stellar rotation. The KOI system shows that the orbital axis is aligned with the stellar spin axis. This system has four planets and dual transits by two of the four planets that were observed during which the two planets overlapped to produce a planet-planet eclipse. Further monitoring of this system will elucidate the details of this rare planetary system.
 
Narita \etal\space reported the first detection of the retrograde orbit of a transiting exoplanet, HAT-P-7b, from their RM effect analysis~\cite{Narita2009}. The further detection of a backward-spinning star with two coplanar planets is also reported~\cite{Hjorth2021}. These findings add constraints on the formation scenario of planetary systems.

\subsection{\bf{The Solar System}}
\begin{figure}[h]
\centering
\includegraphics[scale=0.45]{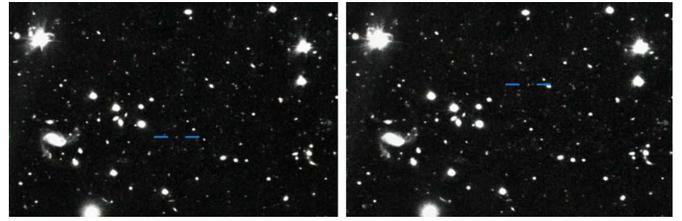}
\vspace{1mm} \caption{(Color online) Images taken on January 15, 2018 (left) and January 16, 2018 (right) show the moving object 2018 AG$_{37}$. The figures are edited from those posted at Institute for Astronomy, University of Hawaii web site~\cite{Sheppard2021}.}
\label{Farfarout2}
\end{figure}

Sheppard \etal\space reported on the finding the most distant planetoid, 2018 AG$_{37}$,  tentatively named as {\lq}{\lq}Farfarout'', at a distance 132 au (Figure \ref{Farfarout2})~\cite{Sheppard2021}. Its elongated orbit reaches as far as 175 au with a pericenter at 27 au. A close gravitational interaction in the past with Neptune is inferred. The size of {\lq}{\lq}farfarout'' is estimated to be at 400 km across from its reflective brightness.
{\lq}{\lq}Farfarout'' is even more remote than the previous solar system distance record-holder, 2018 VG$_{18}$, nicknamed as {\lq}{\lq}Farout{\rq}{\rq} by the same group. These objects offer a clue to search for the undetected Planet Nine.

Batygin \etal\space pointed out that the observed physical clustering of orbits with a semi-major axis exceeding $\sim$250 au, the detachment of perihelia of select Kuiper belt objects from Neptune and the dynamical origin of highly inclined/retrograde long-period orbits are elusive of the existence of Planet Nine~\cite{Batygin2019}. Searching for additional objects, like 2018 AG$_{37}$, is feasible only with wide-field cameras, like HSC.

%\bf{~\citet{Sugita2005} made COMICS imaging observation of the comet 9P/Tempel 1 before and after its collision with the 370-kg Deep Impact probe maneuvered by NASA. Mid infrared observation confirmed a large mass of dust, 10$^{6}$ kg, was ejected forming two wings consistent with the gravitational dynamical simulations.  The dust distribution in the inner part of the plume was found inconsistent with a pure gravitational control and implies that evaporation and expansion of volatiles accelerated the dust.} \rm

\vspace{5mm}
\section{\bf{20 Years of Operation}}
\subsection{\bf{Policy and Statistics}}
After the first light in January 1999, it took 1.5 years to tune the performance of the Subaru Telescope and to commission some of the observational instruments. Eventually, the Subaru Telescope was made available for open use since December 2000.  To obtain open-use telescope time, astronomers must submit an observing proposal with a scientific rationale and feasibility of the proposal.  Several international experts are asked to review and evaluate each proposal. The evaluation reports are assembled to choose winners by the Subaru Time Allocation Committee (TAC) under the Science Advisory Committee (SAC). The oversubscription rate is between 3 and 5. TAC took the initial policy for two years to allocate observing time up to five nights for each program to encourage potential users to get familiar with the facility. To facilitate observational programs that require a larger number of nights, the SAC prepared the Intensive Program framework (since 2003) and the SSP framework (since 2007) for more systematic observational programs.

So far, four SSP programs have been approved: (1) SEED SSP (2009--15), ``Strategic Explorations of Exoplanets and Disks with Subaru"~\cite{Tamura2016}, made 120 nights of observation of exoplanets and proto-planetary disks using the stellar coronagraph instruments, CIAO and HiCIAO, behind the AO188. (2) FastSound (2011--14), ``A Near-Infrared Galaxy Redshift Survey for Cosmology at $z\sim1.3$ using Subaru/FMOS"~\cite{Tonegawa2015}, finished 40 nights of observations using FMOS. (3) HSC-SSP (2014--21), ``Wide-field imaging with Hyper Suprime-Cam: Cosmology and Galaxy Evolution"~\cite{Aihara2018a}, is about to complete 330 nights observation. (4) IRD-SSP (2019--25), ``Search for Planets like Earth around Late-M Dwarfs: Precise Radial Velocity Survey with IRD"~\cite{Kotani2020}, is under way to conduct 175 nights of observation.

Currently, on average, observing nights are allocated to Open Use (49\%), SSP (17\%), University of Hawaii (15\%), Engineering (10\%) and Director's Discretion Time (10\%), respectively. Remote and service observations are gradually increased to ensure a more efficient operation. The implementation of unmanned summit operation is planned on a limited basis.

The observed data are preserved to the proposers of the respective observing programs as proprietary data for 18 months. After this proprietary period, data are made available to any registered researcher in the world through the data archive system. For most of the data, a system SMOKA system is available~\cite{Baba2002}. Additionally, for instance, the photometric redshifts for HSC-SSP imaging data are available from the dedicated site~\cite{Tanaka2018}.

Subaru Telescope experienced two major earthquakes. The telescope tracking was found to have a significantly larger error right after the operation was resumed after a magnitude 6.7 earthquake occurred on October 15, 2006. This was caused by a 1.3 mm displacement of the telescope position on its pier. A correction of the coordinates of the telescope resolved the issue. Another magnitude 6.9 earthquake hit the telescope on May 4, 2018 causing partial damage to the enclosure.

\begin{figure}[h]
\centering
\includegraphics[scale=0.45]{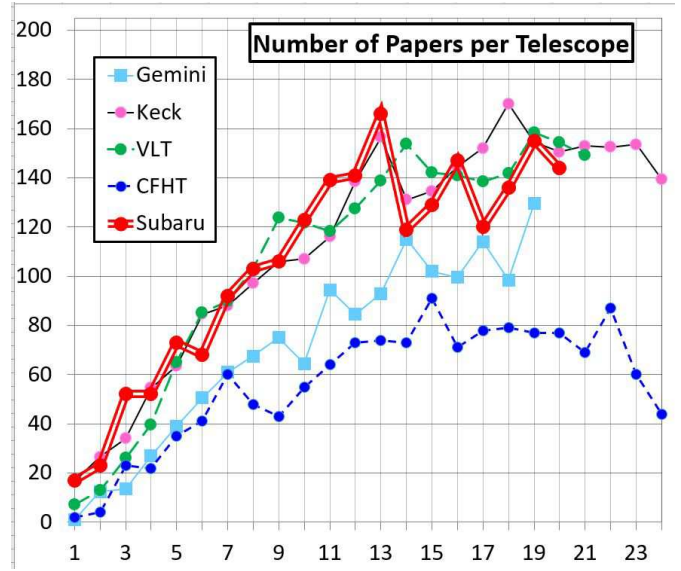}
\vspace{1mm} \caption{(Color online) Comparison of the growth in the number of published  papers per telescope as a function of the year from commissioning. Subaru telescope (red double line) is competing with Keck (pink solid line) and VLT (green broken line) and exceeds Gemini (light blue line with square symbols) and CFHT (blue broken line).  The figure was redrawn from the original data~ \cite{Yoshida2019}.}
\label{paper-rate}
\end{figure}

Figure \ref{paper-rate} shows the number growth curves of annual refereed papers of 8--10 m telescopes. This graph shows the scientific competitiveness of Subaru Telescope.

%\begin{figure}[h]
%\centering
%\includegraphics[scale=0.25]{TimeAssignment.eps}
%\vspace{1mm} \caption{The observing time assignment for 2019. Director's discretion time (DDT), Engineering time(Eng), University of Hawaii time (UH),  Subaru Strategic Program time (SSPs) and Open Use time.}
%\label{TimeAssignment}
%\end{figure}

\subsection{\bf{Future Plans}}
Observational astronomy has made giant leaps during the last several decades. Many space missions have enlarged the wavelength coverage and the survey volume coverage in observational astronomy.
Neutrino and gravitational wave detection became possible and their connection to observational astronomy is referred to as multi-messenger astronomy.   Subaru Telescope started as an all-mighty telescope, but concentrating on its unique wide-field survey capability will be the main role in the coming decades. 

Hattori \etal\space summarized the plan of Subaru instrumentation as of December 2020~\cite{Hattori2020}. The past two decades achievements have revealed the advantage of the unique wide-field capability of the Subaru Telescope. The plan, therefore, places an emphasis on using the currently available HSC for wide-field imaging survey programs. 

The Prime Focus Spectrograph (PFS), which is currently under commissioning and will start scientific observations in 2023, has been developed based on an international collaboration led by Kavli-IPMU and NAOJ. The consortium includes universities and institutes of USA, UK, France, Germany, Taiwan, and Brazil.
PFS enables the spectroscopy of 2400 sources simultaneously for a spectral region 0.38--1.26 $\mu$m with its multifiber feed mechanism~\cite{Tamura2018}.  The combination of HSC and PFS will be a powerful tool for the coming decade of the Subaru Telescope. 
Takada \etal\space have summarized the science goals, using the planned PFS survey: the history and fate of the Universe and the physical process and role of dark matter in governing the assembly of galaxies over the cosmic time including the Milky Way Galaxy~\cite{Takada2014}. 

The third wide-field instrument under R\&D study, ULTIMATE, is an imager and a multi-object spectrograph covering a 20 arcmin field of view for 1.0--2.5 $\mu$m with the help of ground layer adaptive optics~\cite{Minowa2020}.

Strategic observing time exchanges of the Subaru Telescope with Keck Telescopes and Gemini Telescope have been arranged to benefit community needs on both sides.  Future scientific collaborations with the Large Synoptic Survey Telescope (LSST) of Vera C. Rubin Observatory, under construction in the southern hemisphere, and astronomical space missions to be launched soon, including the 1.2 m Euclid mission, the 2.4 m Rowan Space Telescope and the 6.5m James Web Space Telescope, are under discussion.

\subsection{\bf{Maunakea and Hawaii Culture}}
The TMT International Observatory, of which the NAOJ is a founding member, plans to build a next-generation telescope with a segmented primary mirror of 30 m in diameter. The TMT with its 4-times higher spatial resolution and 13-times larger light-collecting power enables a point-source detection sensitivity 180-times more than Subaru Telescope. As Subaru Telescope remains the only telescope in the northern hemisphere capable of making deep and wide survey observations, it will bring forward important candidates of observations for TMT. Thus, the combination of Subaru Telescope and TMT, fully equipped with adaptive optics, has been expected to show new visions of the Universe.

\begin{figure}[h]
\centering
\includegraphics[scale=0.45]{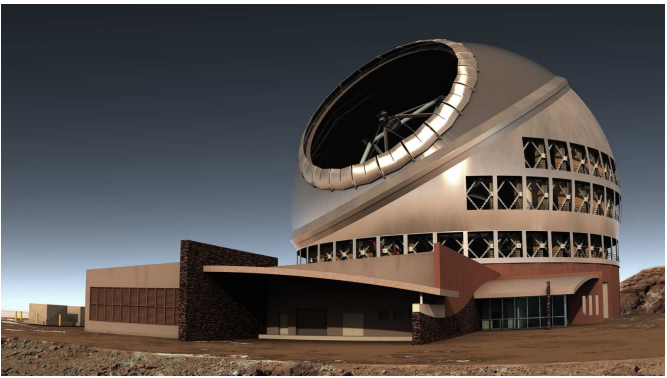}
\vspace{1mm} \caption{(Color online) CG rendering of the TMT to be built on Maunakea (TMT International Observatory).}
\label{TMT}
\end{figure}

The Maunakea site of Subaru Telescope has a land sublease agreement with University of Hawaii valid until 2033. The University of Hawaii is preparing an official process to apply for an extension of the master lease from the State of Hawaii.

When TMT International Observatory planned to have a groundbreaking event at the summit of Maunakea in October 2014, the access road near the summit was blocked by hundreds of protestors. TMT was chosen as a visible target of protest movement against the historical injustice to the native Hawaiian community. Several lawsuits and illegal blockings of the access road suspended the start of the construction work. Due to the sensitive nature of the issue, the local authorities have not been able to resolve this deadlock situation. 

Subaru Telescope and the TMT teams expanded much effort to establish a healthy relationship with the local community. The history of ancient Hawaii began with Polynesians and Micronesians navigating canoes to travel across the ocean and reaching the Hawaii islands of around AD 400. Nainoa Thompson has shown that navigating canoes between islands was possible without modern instruments. Careful observations of stars, winds, and waves enabled them to navigate among the islands. He called the method wayfinding or star navigation~\cite{Low2007}. In this sense, we understand that the native Hawaiians culture and modern astronomy have common roots.

Although this cultural issue is beyond the control of the astronomical community, it is vital for the astronomy community to find a way to continue astronomical observations on Maunakea toward the future. Science and culture are not mutually exclusive and can coexist. This author wishes everyone involved in the future of Maunakea to find an agreeable best solution, and to sustain the availability of this precious astronomy site for the benefit of all humankind.

\vspace{5mm}
\section{\bf{Conclusion} \label{sec:discussion}}
This paper records the history of Subaru Telescope from its beginning in the early 1980s until 2020 through the author's perspective, while covering the engineering aspects and some of the scientific results that could be achieved through the evolving telescope. 

The key engineering of Subaru Telescope was the implementation of technology to enable active control of the optical shape of the primary mirror so as to provide the best image quality under varying gravitational and thermal conditions. Another important development of the adaptive optics system with a laser-guide-star facility made a 10 times improvement of the image quality by cancelling out the atmospheric disturbance to the image in real time.

Various scientific discoveries and achievements made in observational cosmology and in studies of exoplanets are highlighted in this review. There are, however, tremendous collections of scientific papers produced through the observations of Subaru Telescope, and some of the selected papers are also introduced in this review.

The capability of wide-field survey sciences with Suprime-Cam, HSC, and PFS established the unique roles of Subaru Telescope in the era of TMT in the northern hemisphere. This is shown by the fact that HSC and PFS gathered strong interest from the international astronomical community. The needs for identifying relevant observational targets in the northern sky from deep and wide-field surveys are other rationales to support the accomplishment of Subaru Telescope for TMT.

 The Japanese astronomical community made a giant leap through the Subaru Telescope  project to reach the top level in world optical/infrared observational sciences. NAOJ gained experiences to conduct this level of major scientific exploration on an international platform. The final notes show some issues that the future astronomy community needs to face.

\vspace{5mm}
\acknowledgments
The project design, construction, and operation of Subaru Telescope have been possible only with the dedications of thousands of individuals from NAOJ, universities, involving industries, governmental entities, and communities, not limited to Japan,  for over the four decades. Their contributions in engineering, sciences, administration, and general support, are all indispensable. The author humbly acknowledges their valuable contributions. The financial support from the Ministry of Education, Culture, Sports, Science and Technology (MEXT) for the construction and operation of Subaru Telescope has been the key to success of Subaru Telescope and is greatly appreciated.

The choices of the scientific achievements included in this review are inevitably biased to the taste of the author. He sincerely apologizes to those whom he has failed to cite explicitly for their contributions for the project and for scientific information. 

The author is grateful especially to Prof. Richard Ellis for his careful reading and constructive comments as a referee and to Dr. Masafumi Yagi for his numerous helpful comments. Dr. Fred Myers gave many suggestions to improve language. He thanks another anonymous referee for useful comments and thanks Drs. W. Aoki, I. Iwata, K. Nariai, T. Sasaki, M. Takada, W. Tanaka, and M. Yoshida for their helpful comments in the early draft of this paper. The author acknowledges Ms. C. Yoshida for her compilation of records on published papers.

We are honored and grateful for the Hawaiian community providing us the opportunity of observing the Universe from Maunakea, having the superb cultural, historical, and natural significance. We hope Maunakea remains the symbol for the coexistence of science and culture.

\newpage
\bibliographystyle{article}

\end{document}